\documentclass[runningheads]{llncs}

\usepackage[ruled,vlined]{algorithm2e}
\usepackage{array}
\usepackage{amsfonts}
\usepackage{amsmath}
\usepackage{amssymb}
\usepackage{caption}
\usepackage{color}
\usepackage{enumerate}
\usepackage[mathscr]{eucal}
\usepackage[in]{fullpage}
\usepackage[toc,section=section,numberedsection ]{glossaries}
\usepackage{graphicx,epstopdf}
\usepackage{hyperref}
\usepackage{latexsym}
\usepackage{multirow}
\usepackage[numbers]{natbib}
\usepackage{newtxmath}
\usepackage{newtxtext}
\usepackage{srcltx} 
\usepackage{subcaption}
\usepackage{times}
\usepackage{url}
\usepackage{verbatim}
\usepackage{xspace}
\usepackage{wrapfig}
\usepackage[table]{xcolor} 

\graphicspath{ {figures/} }

\newcommand{\eg}{\emph{e.g.},\xspace}

\newcommand{\ie}{\emph{i.e.,}\xspace}

\newcommand{\etal}{\emph{et~al.}\xspace}
\newcolumntype{x}[1]{>{\raggedright\arraybackslash}p{#1}}

\SetCommentSty{mycommfont}

\DeclareMathOperator*{\argmax}{arg\,max}

\makeatletter 
\g@addto@macro{\@algocf@init}{\SetKwInOut{Output}{Output}} 
\makeatother

\begin{document}

\title{Reinforcement Learning for Autonomous Defence in Software-Defined Networking
}

\author{Yi Han\inst{1}\orcidID{0000-0001-6530-4564} \and 
Benjamin I.P. Rubinstein\inst{1}\orcidID{0000-0002-2947-6980} \and 
Tamas Abraham\inst{2}\orcidID{0000-0003-2466-7646} \and 
Tansu Alpcan\inst{1}\orcidID{0000-0002-7434-3239} \and 
Olivier De Vel\inst{2} \and  
Sarah Erfani\inst{1}\orcidID{0000-0003-0885-0643} \and 
David Hubczenko\inst{2} \and 
Christopher Leckie\inst{1}\orcidID{0000-0002-4388-0517} \and 
Paul Montague\inst{2}
}

\authorrunning{Y. Han et al.}

\institute{School of Computing and Information Systems, The University of Melbourne\\ \and 
Defence Science and Technology Group \\
\email{\{yi.han, benjamin.rubinstein, tansu.alpcan, sarah.erfani, caleckie\}@unimelb.edu.au \\
\{tamas.abraham, olivier.devel, david.hubczenko, paul.montague\}@dst.defence.gov.au}}

\maketitle

\begin{abstract}
Despite the successful application of machine learning (ML) in a wide range of domains, adaptability---the very property 
that makes machine learning desirable---can be exploited by adversaries to contaminate training and evade classification. 
In this paper, we investigate the feasibility of applying a specific class of machine learning algorithms, namely, 
reinforcement learning (RL) algorithms, for autonomous cyber defence in software-defined networking (SDN). 
In particular, we focus on how an RL agent reacts towards different forms of causative attacks that poison its training process, 
including indiscriminate and targeted, white-box and black-box attacks. In addition, we also study the impact of the attack timing, 
and explore potential countermeasures such as adversarial training.

\keywords{Adversarial reinforcement learning, Software-Defined Networking, cyber security, adversarial training.}
\end{abstract}

\section{Introduction}\label{sec:intro}
Machine learning has enjoyed substantial impact on a wide range of applications, from cyber-security 
(\eg network security operations, malware analysis) to 
autonomous systems (\eg decision-making and control systems, computer vision). Despite the many successes, the very property 
that makes machine learning desirable---adaptability---is a vulnerability to be exploited by an economic competitor or
state-sponsored attacker. Attackers who are aware of the ML techniques being deployed can contaminate the training data to 
manipulate a learned ML classifier in order to evade subsequent classification, or can manipulate the metadata 
upon which the ML algorithms make their decisions and exploit identified weaknesses in these algorithm---so called 
Adversarial Machine Learning~\cite{barreno_security_2010,huang2011adversarial,biggio_poisoning_2012}.

This paper focuses on a specific class of ML algorithms, namely, reinforcement learning (RL) algorithms, and 
investigates the feasibility of applying RL for autonomous defence in computer networks~\cite{beaudoin_autonomic_2009}, 
\ie the ability to ``fight through'' a contested environment---in particular adversarial machine learning attacks---and ensure 
critical services (\eg email servers, file servers, etc.) are preserved to the fullest extent possible. 

\begin{figure}
\centering
\includegraphics[width=.67\textwidth]{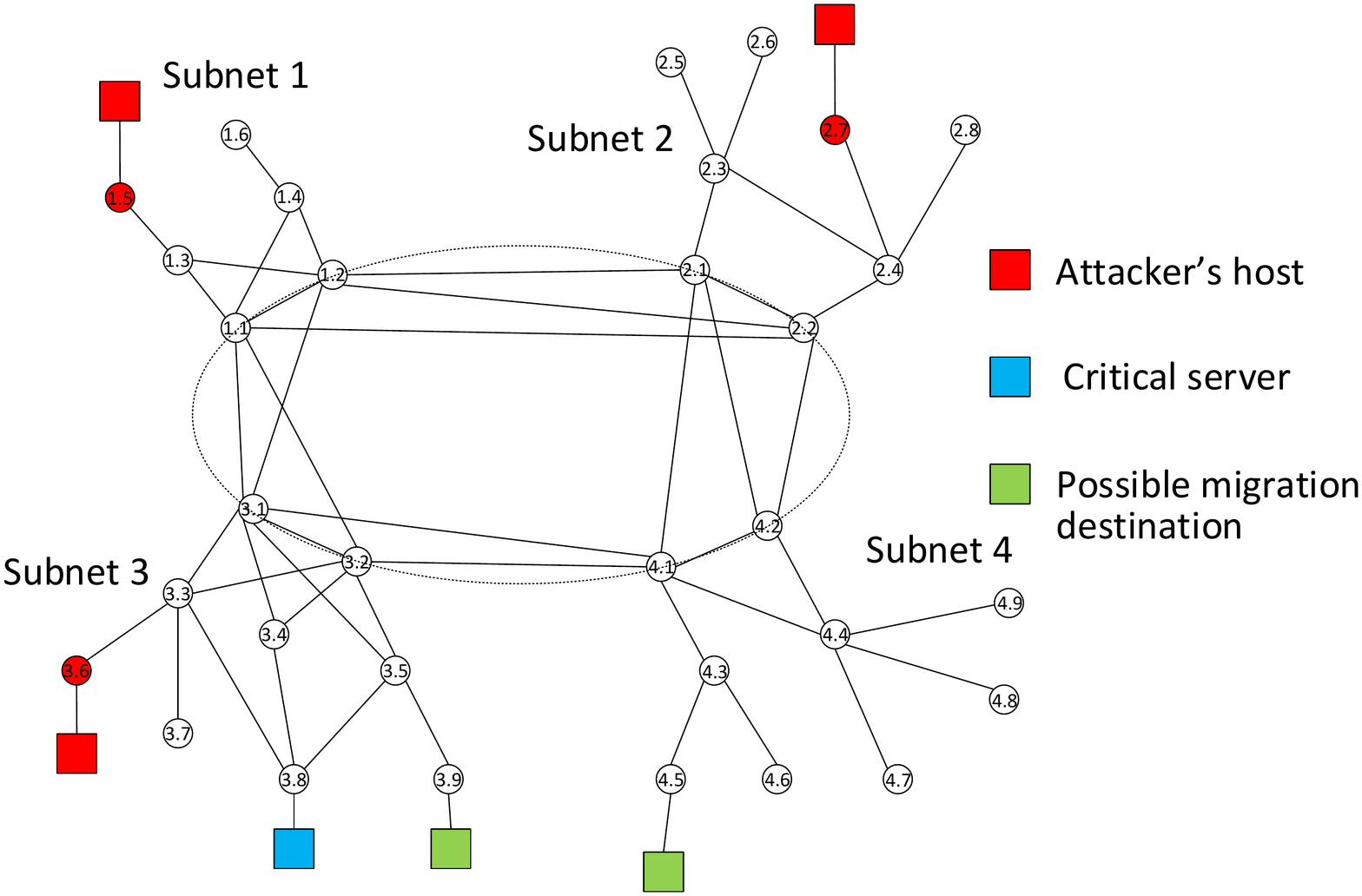}
\caption{An example network setup. The attacker propagates through the network to compromise the critical server, 
and the defender applies RL to prevent the critical server from compromise and to preserve as many nodes as possible.}
\label{figure_platform}
\end{figure}

For example, consider a network as shown in Fig.~\ref{figure_platform} that consists of 32 nodes, one (node 3.8) of whom connects to 
the critical server, two (nodes 3.9 and 4.5) connect to potential migration destinations, and three (nodes 1.5, 2.7 and 3.6) 
connect to the attacker's hosts. The attacker 
aims to propagate through the network, and compromise the critical server. We aim to prevent this and preserve as many nodes 
as possible through the following RL approach: 
\begin{itemize}
\item We first train two different types of RL agents: Double Deep Q-Networks (DDQN)~\cite{van_hasselt_deep_2015} and 
Asynchronous Advantage Actor-Critic (A3C)~\cite{mnih_asynchronous_2016}. 
The agents observe network states, and select actions such as ``isolate'', ``patch'', 
``reconnect'', and ``migrate''. The agents gradually optimise their actions for different network states, based on the received 
rewards for maintaining critical services, costs incurred when shutting down non-critical services or migrating critical 
services. 
\item Once a working agent is obtained, we then investigate different ways by which the attacker may poison the training 
process of the RL agent. For example, the attacker can falsify part of the reward signals, or manipulate the states of 
certain nodes, in order to trick the agent to take non-optimal actions, resulting in either the critical server being 
compromised, or significantly fewer nodes being preserved. Both indiscriminate (untargeted) and targeted, white-box and black-box 
attacks are studied. 
\item We also explore possible countermeasures---\eg adversarial training---that make the training less vulnerable to 
causative/poisoning attacks.
\item To make use of the developed capacity for autonomous cyber-security operations, we build our experimental platform around 
software-defined networking (SDN)~\cite{noauthor_sdn_2014}, a next-generation tool chain for centralising and abstracting 
control of reconfigurable networks. The SDN controller provides a centralised view of the whole network, and is directly 
programmable. As a result, it is very flexible for managing and reconfiguring various types of network resources. 
Therefore, in our experiments the RL agents obtain all network information and perform different network operations 
via the SDN controller.
\item Our results demonstrate that RL agents can successfully identify the optimal actions to protect the critical server, 
by isolating as few compromised nodes as possible. In addition, even though the causative attacks can cause the agent to 
make incorrect decisions, adversarial training shows great potential for mitigating the negative impact.
\end{itemize}

The remainder of the paper is organised as follows: 
Section~\ref{sec:preliminaries} briefly introduces the fundamental concepts in reinforcement learning and SDN; 
Section~\ref{sec:problem} defines the research problem; 
Section~\ref{sec:attack} introduces in detail the different forms of proposed attacks against RL; 
Section~\ref{sec:experiment} presents the experimental results on applying RL for autonomous defence in SDN, and the impact of 
those causative attacks; 
Section~\ref{sec:related} overviews previous work on adversarial machine learning (including attacks against reinforcement 
learning) and existing countermeasures; 
Section~\ref{sec:conc} concludes the paper, and offers directions for future work. 

\section{Preliminaries}\label{sec:preliminaries}
Before defining the research problems investigated in this paper, we first briefly introduce the basic concepts in 
reinforcement learning and software-defined networking.

\subsection{Reinforcement Learning}
In a typical reinforcement learning setting~\cite{sutton_introduction_1998}, an agent repeatedly interacts with the environment: 
at each time step \(t\), the agent (1) observes a state \(s_{t}\) of the environment; (2) chooses an action \(a_{t}\) 
based on its policy \(\pi\)---a mapping from the observed states to the actions to be taken; and (3) receives 
a reward \(r_{t}\) and observes next state \(s_{t+1}\). This process continues until a terminal state is reached, 
and then a new episode restarts from a certain initial state. 

The agent's objective is to maximise its discounted cumulative rewards over the long run:  
\(R_{t} = \sum_{\tau=t}^{\infty} \gamma^{\tau-t}r_{\tau}\), where \(\gamma \in (0, 1]\) is the discount factor that controls the 
trade-off between short-term and long-term rewards.

Under a given policy \(\pi\), the value of taking action \(a\) in state \(s\) is defined as: 
\(Q^{\pi}(s, a) = \mathop{\mathbb{E}}[R_{t}|s_{t}=s, a_{t}=a, \pi]\). Similarly, the value of state \(s\) is defined as: 
\(V^{\pi}(s) = \mathop{\mathbb{E}}[R_{t}|s_{t}=s, \pi]\). In this paper, we mainly focus on two widely cited RL algorithms: Double Deep Q-Networks 
(DDQN)~\cite{van_hasselt_deep_2015} and Asynchronous Advantage Actor-Critic (A3C)~\cite{mnih_asynchronous_2016}.

\subsubsection{Q-Learning}
Q-learning~\cite{sutton_introduction_1998} approaches the above problem by estimating the optimal action value function 
\(Q^{*}(s, a) = \max_{\pi} Q^{\pi}(s, a)\). Specifically, it uses the Bellman equation 
\(Q^{*}(s, a) = \mathop{\mathbb{E}}_{s^{\prime}}[r+\gamma \max_{a^{\prime}} Q^{*}(s^{\prime}, a^{\prime})]\) to update the value
iteratively. In practice, Q-learning is commonly implemented by function approximation with parameters \(\theta\): 
\(Q^{*}(s, a) \approx Q(s, a; \theta)\). At each training iteration \(i\), the loss function is defined as: 
\(L_{i}(\theta_{i}) = \mathop{\mathbb{E}}[(r+\gamma \max_{a^{\prime}} Q(s^{\prime},a^{\prime};\theta_{i-1}) - 
Q(s,a;\theta_{i}))^{2}]\).

\subsubsection{Deep Q-Networks (DQN)}
Classic Q-learning networks suffer from a number of drawbacks, including (1) the \textit{i.i.d.} (independent and identically distributed) 
requirement of the training data being violated as consecutive observations are correlated, 
(2) unstable target function when calculating Temporal Difference (TD) errors, 
and (3) different scales of rewards. Deep Q networks (DQN)~\cite{mnih_playing_2013} overcome these issues by 
(1) introducing experience replay, (2) using a target network that fixes its parameters (\(\theta^{-}\)) 
and only updates at regular intervals, and (3) clipping the rewards to the range of \([-1, 1]\). 
The loss function for DQN becomes: \(L_{i}(\theta_{i}) = \mathop{\mathbb{E}}[(r+\gamma \max_{a^{\prime}} 
Q(s^{\prime},a^{\prime};\theta^{-}_{i}) - Q(s,a;\theta_{i}))^{2}]\).

\subsubsection{Double DQN (DDQN)} To further solve the problem of value overestimation, Hasselt \etal~\cite{van_hasselt_deep_2015} 
generalise the Double Q-learning algorithm~\cite{hasselt_double_2010} proposed in the tabular setting, and propose 
Double DQN (DDQN) that separates action selection and action evaluation, \ie one DQN is used to determine the maximising action 
and a second one is used to estimate its value. Therefore, the loss function is: 
\(L_{i}(\theta_{i}) = \mathop{\mathbb{E}}[(r+\gamma Q(s^{\prime},\argmax_{a^{\prime}} Q(s^{\prime},a^{\prime};\theta_{i});\theta^{-}_{i}) 
- Q(s,a;\theta_{i}))^{2}]\).

\subsubsection{Prioritised Experience Replay}
Experience replay keeps a buffer of past experiences, and for each training iteration, it samples uniformly a batch of experiences 
from the buffer, \ie each experience is treated equally regardless of importance. 
Prioritised experience replay~\cite{schaul_prioritized_2015} assigns 
higher sampling probability to transitions that do not fit well with the current estimation of the \(Q\) function. For DDQN, the error 
of an experience is defined as \(|r+\gamma Q(s^{\prime},\argmax_{a^{\prime}} Q(s^{\prime},a^{\prime};\theta);\theta^{-}) 
- Q(s,a;\theta)|\). Experiences with larger errors are more likely to be chosen, and the errors will be updated at the end of 
each training iteration.

\subsubsection{Asynchronous Advantage Actor-Critic (A3C)}
Mnih \etal~\cite{mnih_asynchronous_2016} propose an asynchronous variant of the classical actor-critic algorithm, which 
estimates both the state value function \(V(s; \theta_{v})\) and a policy \(\pi(a|s; \theta_{p})\). 
Specifically, the asynchronous advantage actor-critic (A3C) algorithm uses multiple threads to explore different 
parts of the state space simultaneously, and updates the global network in an asynchronous way.
In addition, instead of using discounted returns to determine whether an action is good, A3C estimates the \emph{advantage function} 
so that it can better focus on where the predictions are lacking.

\subsection{Software-Defined Networking}
In conventional online applications, the majority of the communication occurs directly between the client and the server. 
However, due to the wide use of content-delivery networks (CDN) and cloud services, it has become common 
for modern applications to access multiple servers and databases before data is finally returned to the client, 
resulting in a significantly higher proportion of ``east-west'' traffic. In addition, users nowadays may access an application 
with any type of device, \eg smartphones, tablets and laptops, from anywhere around the world. 
These changes have contributed to an ever-increasing strain on traditional networks, largely due to conventional switches 
having their own control unit, making network reconfiguration time consuming.

In order to better serve today's dynamic and high-bandwidth applications, a new architecture called 
Software-Defined Networking (SDN) has emerged~\cite{noauthor_sdn_2014}. There are three layers in the SDN architecture: 
(1) the application layer includes applications that deliver services. These applications communicate their network requirements 
to the controller via northbound APIs; (2) the SDN controller translates these requirements into low-level controls, 
and sends them through southbound APIs to the infrastructure layer; (3) the infrastructure layer comprises network switches 
that control forwarding and data processing.

One major advantage of SDN is that it decouples network control and forwarding functions, rendering the controller 
directly programmable. As a result, network resources can be conveniently managed, configured and optimised using 
standardised protocols. For example, Openflow~\cite{noauthor_openflow_2011} is a commonly-used protocol between 
the controller and the underlying switches. Note that in the simplest implementation, the controller follows a 
centralised design where it has a global view of the entire network. However, in order to manage large-scale networks, 
hierarchical and distributed designs can also be adopted.

There have been a number of proprietary and open-source SDN controller software platforms, 
such as Cisco's Open SDN Controller~\cite{noauthor_cisco_2017}, Floodlight~\cite{noauthor_floodlight_2017}, 
NOX/POX~\cite{noauthor_nox/pox_2017} and Open vSwitch~\cite{noauthor_open_2016}. 
In this paper, we have opted to use OpenDaylight~\cite{medved_opendaylight:_2014,noauthor_opendaylight_2017}, 
which is the largest open-source SDN controller today and which is updated regularly.

\subsubsection{Applying RL in SDN}\label{sec_rl_sdn}
One main challenge SDN faces arises from highly-dynamic traffic patterns, which motivate a requirement for the network to be 
reconfigured frequently. It has been demonstrated that RL is an ideal tool to accomplish such a task
~\cite{salahuddin_software-defined_2015,mao_resource_2016,lin_qos-aware_2016,huang_energy-efficient_2015,wang_novel_2016,
salahuddin_reinforcement_2016,mestres_knowledge-defined_2017,duggan_reinforcement_2016,kim_congestion_2016,cao_service_2017}. 
For example, Salahuddin \etal \cite{salahuddin_software-defined_2015} propose a roadside unit (RSU) cloud to enhance traffic 
flow and road safety. This RSU cloud is implemented using SDN, and leverages reinforcement learning to  better cope with the 
dynamic service demands from the transient vehicles. In this way. the reconfiguration overhead can be minimised over the 
long run: increasing/decreasing the number of service centres, migrating services from one centre to another, for example. 
Mao \etal \cite{mao_resource_2016} design a job scheduling algorithm, called DeepRM, for computing clusters that have 
similar demands as SDN management. DeepRM uses colour images to represent system resources, \eg CPU, RAM, as well as the 
resources required by each job slot, and exploits policy gradient methods to minimise average job slowdown.

\section{Problem Statement}\label{sec:problem}
In this paper, we seek to answer the question: Can reinforcement learning be used for autonomous defence in SDN? 
We start with a scenario that does not consider the attacker poisoning the training process, and then investigate 
the impact of adversarial reinforcement learning. While we also briefly discuss potential countermeasures, we largely leave 
defences to future work.

\subsection{Reinforcement Learning Powered Autonomous Defence in SDN}
Consider a network of \(N\) nodes (\eg Fig.~\ref{figure_platform}), \(H=\{h_{1}, h_{2}, ..., h_{N}\}\), 
where \(H_{C} \subset H\) is the set of critical servers to be protected (blue nodes in Fig.~\ref{figure_platform}), 
\(H_{M} \subset H\) is the set of possible migration destinations for \(h \in H_{C}\) (green nodes), 
and \(H_{A} \subset H\) is the set of nodes that have initially been compromised (red nodes).
The attacker aims to propagate through the network, and penetrate the mission critical servers, 
while the defender/SDN controller monitors the system state, and takes appropriate actions 
in order to preserve the critical servers and as many non-critical nodes as possible.

Reflecting suggestions from past work, we consider a defender adopting RL. In this paper, we start with a simplified version, and make the 
following assumptions (Section~\ref{sec:conc} explains how they may be replaced): 
(1) each node (or link) only has two states: compromised/uncompromised (or on/off); 
(2) both the defender and the attacker know the complete network topology; (3) the defender has in place a 
detection system that can achieve a detection rate of 90\%, with no false alarms (before the causative attacks); 
(4) the attacker needs to compromise all nodes on the path (\ie cannot "hop over" nodes).

Given these assumptions, in each step the defender: 
\begin{enumerate}
	\item Observes the state of the network---whether a node is compromised, and whether a link 
is switched on/off, \eg there are 32 nodes and 48 links in Fig.~\ref{figure_platform}, 
so one state contains an array of 80 \(0s/1s\), where \(0\) means the node is uncompromised or the link is switched off, 
and \(1\) means the node is compromised or the link is switched on; 
	\item Takes an action that may include: 
(i) isolating and patching a node; (ii) reconnecting a node and its links; 
(iii) migrating the critical server and selecting the destination; and (iv) taking no action.
Note that, in this scenario, the defender can only take one type of action at a time, and if they decide to isolate/reconnect, 
only one node can be isolated/reconnected at a time. In the example of Fig.~\ref{figure_platform}, 
there are 68 (32+32+3+1) possible actions altogether; 
	\item Receives a reward based on (i) whether the critical servers are compromised; (ii) the number of nodes 
reachable from the critical servers; (iii) the number of compromised nodes; (iv) migration cost; and 
(v) whether the action is valid, \eg it is invalid to isolate a node that has already been isolated, or to migrate a server 
to the current location.
\end{enumerate}

Meanwhile, the attacker carefully chooses the nodes to compromise. For example, in the setting of 
Fig.~\ref{figure_platform}, they infect a node only if it (1) is closer to the ``backbone'' network 
(nodes on the dashed circle); (2) is in the backbone network; or (3) is in the target subnet. 
Table~\ref{table_problem_description} summarises this problem setting.

\begin{table}[h]\small
\centering
\caption{Problem description: RL powered autonomous defence in SDN}
\label{table_problem_description}
\begin{tabular}{ l x{7.2cm} x{7.2cm} }
\hline
& \textbf{Defender} & \textbf{Attacker} \\
\hline
\textbf{State} & (1) Whether each node is compromised; \newline (2) Whether each link is turned on/off. & \\ 
\hline
\textbf{Actions} & (1) Isolate and patch a node; \newline (2) Reconnect a node and its links; \newline 
(3) Migrate the critical server and select the destination; \newline (4) Take no action & 
Compromise a node that satisfies certain conditions, \eg the node (1) is closer to the ``backbone'' network; 
(2) is in the backbone network; or (3) in the target subnet.\\
\hline
\textbf{Goals} & (1) Preserve the critical servers; \newline (2) Keep as many nodes uncompromised and 
reachable from the critical servers as possible. & Compromise the critical servers.\\
\hline
\end{tabular}
\end{table}

Alternatively, if we look from a game perspective, the problem can be defined as a two-player zero-sum 
security game~\cite{alpcan_network_2011} between the attacker (\(A\)) and the defender (\(D\)): 
\(G=\{P, AS^{i}, U^{i}, i\in\{A,D\}\}\), where \(P\) = \{\(A, D\)\}, \(AS\) is the action set, and \(U\) is the utility. 
Specifically, in the problem setting discussed here, while the defender can choose one of the four types of actions, 
the attacker has a pure/deterministic strategy (we will replace this and model the attacker as another 
agent~\cite{pinto_robust_2017} in future work). 
In addition, the defender's utility \(U^{D}\) is the discounted accumulative rewards 
\(R_{t} = \sum_{\tau=t}^{\infty} \gamma^{\tau-t}r_{\tau}\), \(\gamma \in (0, 1]\), 
and since it is a zero-sum game, \(U^{A}=-U^{D}\). Considering tractability of computing equilibria (especially for much larger 
and more complex networks), we are interested in approximating the optimal response for the defender using RL algorithms.

\subsection{Causative Attacks against RL Powered Autonomous Defence System} \label{ps_casative_attacks}
As an online system, the autonomous defence system continues gathering new statistics, and 
keeps training/updating its model. Therefore, it is necessary and crucial to analyse the impact of an adversarial environment, 
where malicious users can manage to falsify either the rewards received by the agent, or the states of certain nodes. 
In other words, this is a form of causative attack that poisons the training process, 
in order for the tampered model to take sub-optimal actions. 
In this paper, we investigate the two forms of attacks below.
\begin{enumerate}
	\item \textbf{Flipping reward signs.} Suppose that without any attack, the agent would learn the following experience 
\((s, a, s^{\prime}, r)\), where \(s\) is the current system state, \(a\) is the action taken by the agent, 
\(s^{\prime}\) is the new state, and \(r\) is the reward. In our scenario, we permit the attacker to flip the sign of a certain 
number of rewards (\eg 5\% of all experiences), and aim to maximise the loss function of the RL agent. 
This is an extreme case of the corrupted reward channel problem~\cite{everitt_reinforcement_2017}, where the reward may be 
corrupted due to sensor errors, hijacks, etc.
	\item \textbf{Manipulating states.} Again, consider the case where the agent learns an experience \((s, a, s^{\prime}, r)\) 
without any attack.	Furthermore, when the system reaches state \(s^{\prime}\), 
the agent takes the next optimal action \(a^{\prime}\). The attacker is then allowed to introduce one false positive (FP) and 
one false negative (FN) reading in \(s^{\prime}\), \ie one uncompromised/compromised node is reported as 
compromised/uncompromised to the defender. As a result, instead of learning \((s, a, s^{\prime}, r)\), 
the agent ends up observing \((s, a, s^{\prime}+\delta, r^{\prime})\) (where \(\delta\) represents the FP and FN readings), 
and consequently may not take \(a^{\prime}\) in the next step.
\end{enumerate}

\section{Attack Mechanisms}\label{sec:attack}
This section explains in detail the mechanisms of the attacks introduced above.

\subsection{Attack I: Maximise Loss Function by Flipping Reward Signs}
Recall that the DDQN agent aims to minimise the loss function: \(L_{i}(\theta_{i}) = \mathop{\mathbb{E}}
[(r+\gamma Q(s^{\prime},\argmax_{a^{\prime}} Q(s^{\prime},a^{\prime};\theta_{i});\theta^{-}_{i}) - Q(s,a;\theta_{i}))^{2}]\).
In the \(i^{th}\) training iteration, \(\theta_{i}\) is updated according to the gradient of \(\partial L_{i}/\partial\theta_{i}\). 
The main idea for the first form of attack is to falsify certain rewards based on \(\partial L_{i}/\partial r\), 
in order to maximise the loss \(L_{i}\).

Specifically as shown in Algorithm~\ref{algo:att_reward}, after the agent samples a batch of experiences for training, 
we calculate the gradient of \(\partial L_{i}/\partial r\) for each of them, and flip the sign of experience with the 
largest absolute value of the gradient \(|\partial L_{i}/\partial r|\) that satisfies \(r\cdot\partial L_{i}/\partial r < 0\) 
(if \(r\cdot\partial L_{i}/\partial r > 0\) flipping the sign decreases the loss function). 

\begin{algorithm}
\LinesNumbered
\SetKwInOut{Input}{Input}\SetKwInOut{Output}{Output}
\Input{The list of sampled experiences \(L_{E}\); The loss function \(L_{i}\) of the RL agent} 
\Output{The tampered experiences \(L_{E}^{\prime}\)}
\BlankLine

\For{experience \((s_{j}, a_{j}, s^{\prime}_{j}, r_{j})\) in \(L_{E}\)}{
	Calculate \(g = \partial L_{i}/\partial r_{j}\)\;
	\If{\(|g| > maxG\) and \(g\cdot r_{j}<0\) }{
		\(maxG = |g|\)\;
		\(maxIdx = j\)\;
	}
}

\((s_{maxIdx}, a_{maxIdx}, s^{\prime}_{maxIdx}, r_{maxIdx}) \leftarrow (s_{maxIdx}, a_{maxIdx}, s^{\prime}_{maxIdx}, -r_{maxIdx})\);

\Return{\(L_{E}^{\prime}\)}
		
\caption{Attack I -- Flipping reward signs\label{algo:att_reward}}
\end{algorithm}

\subsection{Attack II: Prevent Agent from Taking Optimal/Specific Actions by Manipulating States}
Our experimental results show that the above form of attack is indeed effective in increasing the agent's loss function. 
However, it only delays the agent from learning the optimal actions. Therefore, the second form of attack directly targets
the value function \(Q\) (against DDQN agent) or the policy \(\pi\) (against A3C agent).

\begin{enumerate}
	\item \textbf{Indiscriminate attacks.} For each untampered experience \((s, a, s^{\prime}, r)\), indiscriminate attacks 
falsify the states of two nodes in the new state \(s^{\prime}\), in order to prevent the agent from taking the next optimal action 
\(a^{\prime}\) that has been learned so far (which may be different from the final optimal action for the given state), 
\ie against DDQN agent the attacks minimise \(\max_{a^{\prime}} Q(s^{\prime}+\delta, a^{\prime})\),
while against A3C agent the attacks minimise \(\max_{a^{\prime}} \pi(a^{\prime}|s^{\prime}+\delta)\). 

	\item \textbf{Targeted attacks.} Targeted attacks aim to prevent the agent from taking a specific action (in our case, 
we find that this is more effective than tricking the agent to take a specific action). As an extreme case, this paper allows 
the attacker to know the (final) optimal action \(a^{*}\) that the agent is going to take next (\(a^{*}\) may be different from \(a^{\prime}\)), 
and they seek to minimise the probability of the agent taking that action: for DDQN, the attacks minimise \(Q(s^{\prime}+\delta, a^{*})\); 
for A3C, the attacks minimise \(\pi(a^{*}|s^{\prime}+\delta)\). 

\end{enumerate}

The details of the above two types of attacks are presented in Algorithm~\ref{algo:att_state} (Note: Algorithm~\ref{algo:att_state} 
is for the attacks against DDQN. Due to similarity, the algorithm for attacks against A3C is omitted). In addition, 
we also consider the following variants of the attacks:

\begin{enumerate}
	\item \textbf{White-box attacks vs. Black-box attacks.} In white-box attacks, the attacker can access the model 
under training to select the false positive and false negative nodes, while in black-box attacks, the attacker 
first trains surrogate model(s), and then uses them to choose the FPs and FNs.
	
	\item \textbf{Limit on the choice of FPs and FNs.} The above attacks do not set any limit on the choice of FPs and FNs, 
and hence even though the attacker can only manipulate the states of two nodes each time, overall, 
they still need to be able to control a number of nodes, which is not practical. Therefore, we first run unlimited 
white-box attacks, identify the top two nodes that have been selected most frequently as FPs and FNs respectively, and only allow the 
attacker to manipulate the states of those nodes.
	
	\item \textbf{Limit on the timing of the attack.} The last type of attacks only introduces FPs and FNs in the first \(m\) 
steps (\eg \(m=3\)) in each training episode.

\end{enumerate}

\begin{algorithm}
\LinesNumbered
\SetKwInOut{Input}{Input}
\SetKwInOut{Output}{Output}
\Input{The original experience \((s, a, s^{\prime}, r)\); The list of all nodes \(L_{N}\); 
Target action \(a_{t}\) (\(a_{t}=-1\) for indiscriminate attack); The main DQN \(Q\)}
\Output{The tampered experience \((s, a, s^{\prime}+\delta, r^{\prime})\)}
\BlankLine

\If{\(a_{t} == -1 \)}{
	\tcp{indiscriminate attack}
    \(a_{t} = \argmax_{a^{\prime}} Q(s^{\prime}, a^{\prime})\)\;
	}

\For{node \(n\) in \(L_{N}\)}{
	\If{\(n\) is compromised}{
		mark \(n\) as uncompromised\;
		\If{\(Q(s^{\prime}+\delta, a_{t}) < minQ_{FN}\)}{
			\tcp{\(\delta\) represents the FP and/or FN readings}
			\(FN = n\)\;
			\(minQ_{FN} = Q(s^{\prime}+\delta, a_{t})\)\;
		}
		restore \(n\) as compromised\;
	}
	\Else{
		mark \(n\) as compromised\;
		\If{\(Q(s^{\prime}+\delta, a_{t}) < minQ_{FP}\)}{
			\(FP = n\)\;
			\(minQ_{FP} = Q(s^{\prime}+\delta, a_{t})\)\;
		}
		restore \(n\) as uncompromised\;
	}
}
		
Change node \(FN\) to uncompromised\;

Change node \(FP\) to compromised\;

\Return{\((s, a, s^{\prime}+\delta, r^{\prime})\)}
		
\caption{Attack II -- Manipulating states\label{algo:att_state}}
\end{algorithm}

\section{Experimental Verification}\label{sec:experiment}
This section begins with a discussion of the experimental results obtained when applying RL to autonomous defence in a SDN 
environment without considering causative 
attacks. We then analyse the impact of the two forms of attacks explained in Section~\ref{sec:attack}. Finally, we discuss 
adopting adversarial training as a potential countermeasure, and present some preliminary results. Experiments on causative 
attacks were performed on eight servers (equivalent to two Amazon EC2 t2.large instances and 
six t2.xlarge instances~\cite{noauthor_amazon_nodate}), and each set of experiments was repeated 15 to 25 times.

\subsection{Autonomous Defence in a SDN}
For our experiments, as shown in Fig.~\ref{figure_platform}, we created a network with 32 nodes and 48 links using 
Mininet~\cite{noauthor_mininet:_2017}, one of the most popular network emulators. OpenDaylight
\cite{medved_opendaylight:_2014,noauthor_opendaylight_2017} serves as the controller, 
and monitors the whole-of-network status. The RL agent retrieves network information and takes appropriate operations by calling 
corresponding APIs provided by OpenDaylight. In the setup, the three nodes in red, \ie nodes 1.5, 2.7 and 3.6, have already 
been compromised. Node 3.8 is the critical server to be protected, and it can be migrated to node 3.9 or 4.5.

\begin{figure}[ht!]
\centering
\begin{subfigure}{.5\columnwidth}
  \centering
  \includegraphics[width=\columnwidth]{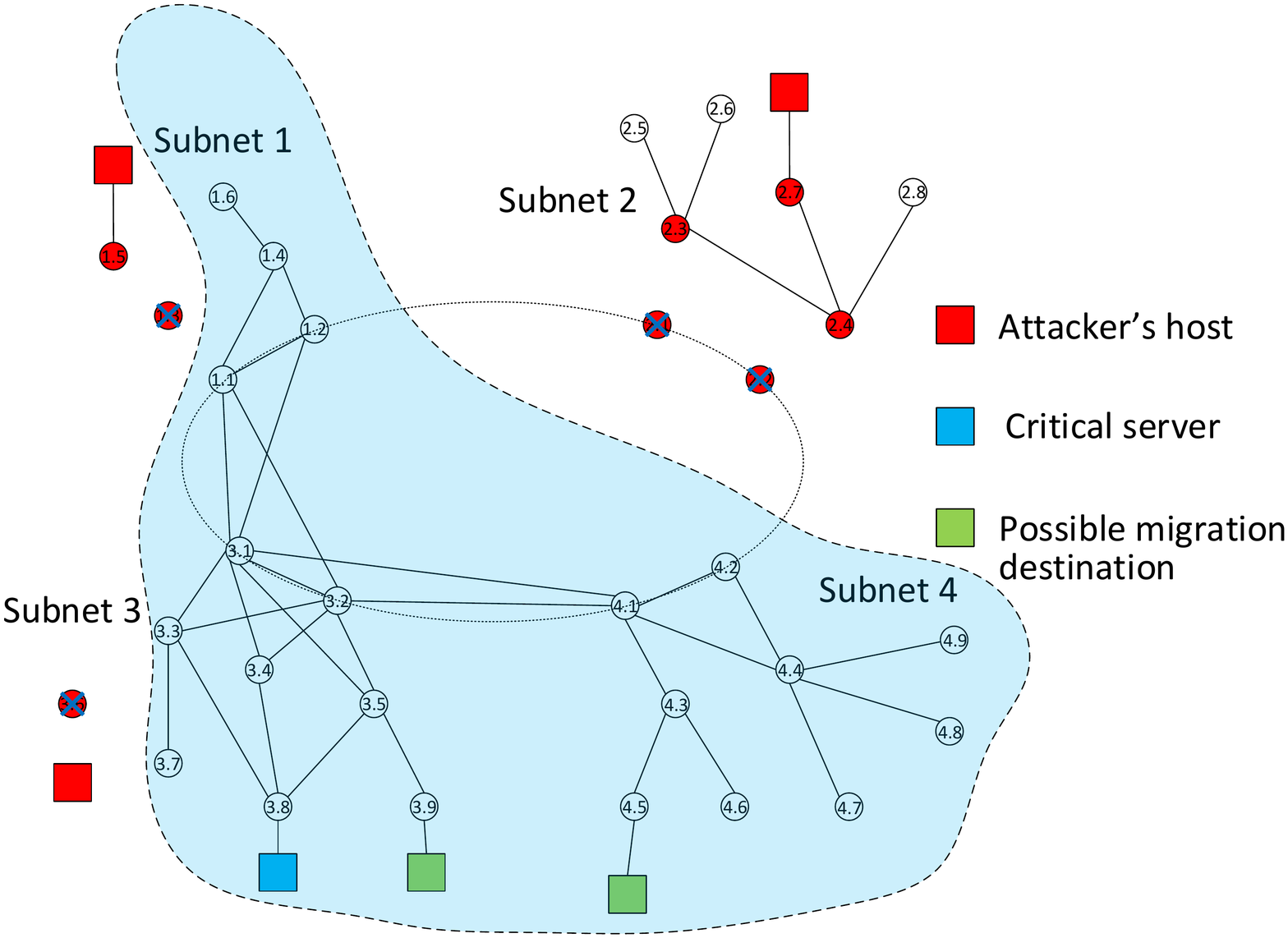}
  \caption{DDQN}
  \label{figure_without_attack_DDQN}
\end{subfigure}%
\begin{subfigure}{.5\columnwidth}
  \centering
  \includegraphics[width=\columnwidth]{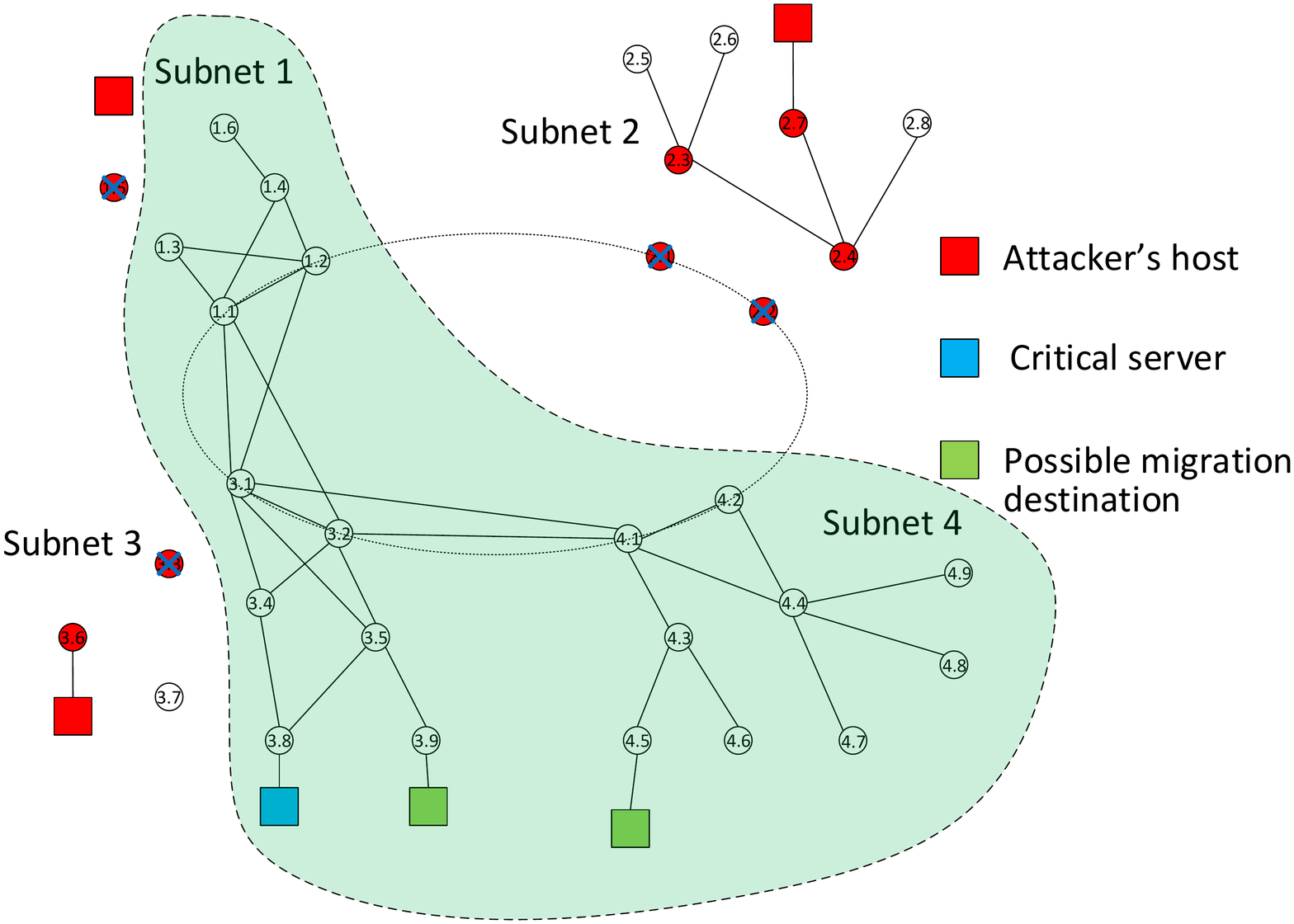}
  \caption{A3C}
  \label{figure_without_attack_A3C}
\end{subfigure}
\caption{Optimal results without causative attacks (nodes in the blue/green shade are preserved)}
\label{figure_without_attack}
\end{figure}

We trained a DDQN (with Prioritised Experience Relay) agent and an A3C agent. We set the length of training such that 
the reward per episode for both agents reached a stable value well before training ended. The two agents learned two slightly 
different responses: the DDQN agent decides to first isolate node 3.6, then 1.3, 2.2 and finally 2.1, 
which means 21 nodes are preserved (see Fig.~\ref{figure_without_attack_DDQN}); 
while the A3C agent isolates nodes 1.5, 3.3, 2.2 and 2.1, 
keeping 20 nodes uncompromised and reachable from the critical server (see Fig.~\ref{figure_without_attack_A3C}).

\subsection{Attack I: Flipping Reward Sign}
This subsection presents the results of the first form of attack that flips the reward sign. In our experiments, we limit the 
total number of tampered experiences to \(5\%\) of all experiences obtained by the agent, and also set the number of tampered 
experiences per training iteration to the range of \([1, 5]\).

\begin{figure}
\centering
\includegraphics[width=.6\textwidth]{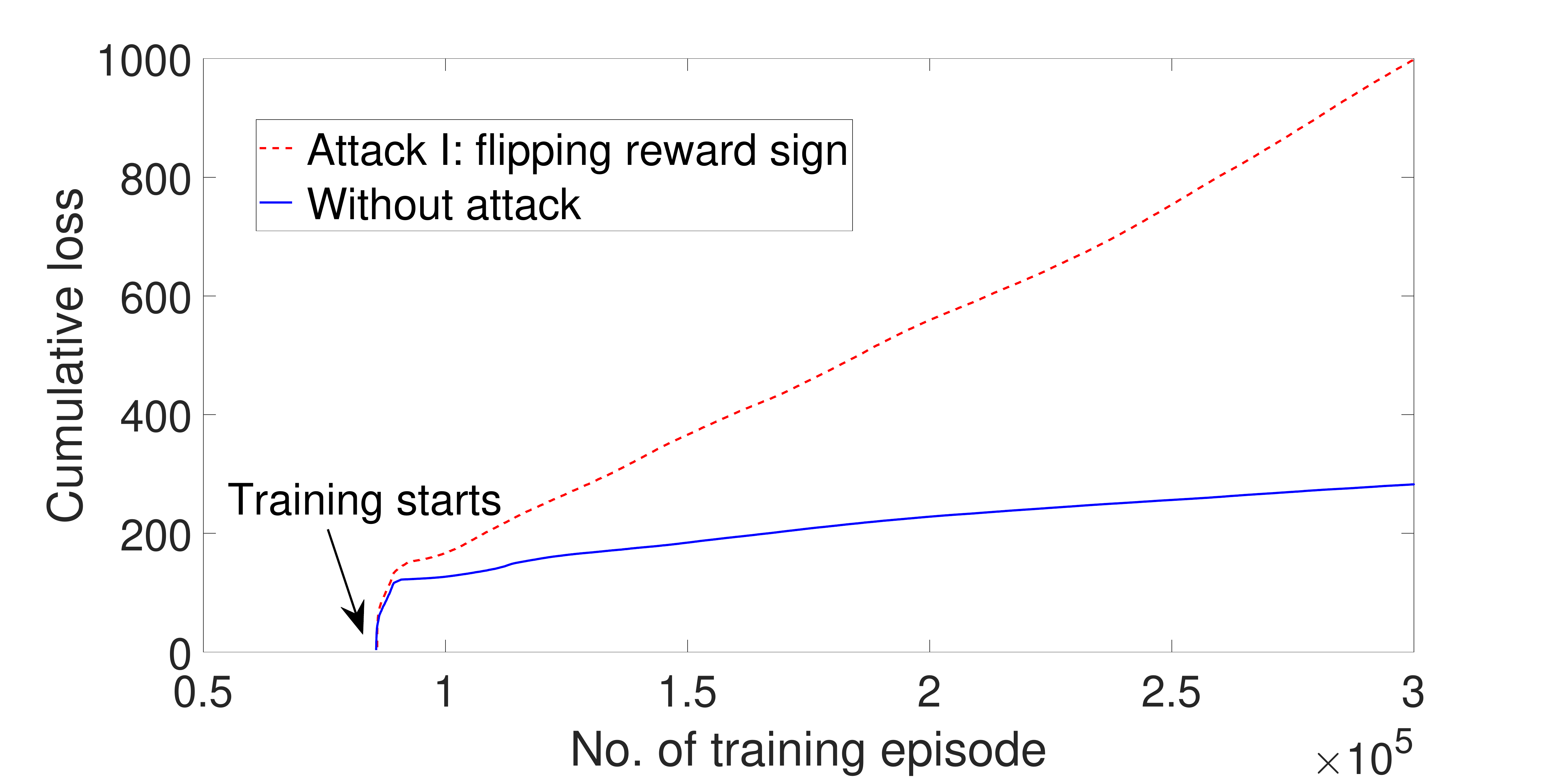}
\caption{Cumulative loss before and after flipping reward sign attacks (against DDQN)}
\label{figure_attack_reward}
\end{figure}

As can be seen in Fig.~\ref{figure_attack_reward}, the attack is effective in increasing the agent's loss function. 
However, our results also suggest that this form of attack only delays the training as the agent still 
learns the optimal actions (although the delay can be significant).

\subsection{Attack II: Manipulate State---Indiscriminate Attacks}

\subsubsection{Unlimited White-Box Attacks}
We start with unlimited indiscriminate white-box attacks, the case where the attacker has full access to the model under training. 
For each experience \((s, a, s^{\prime}, r)\) obtained by the agent, they can manipulate the states of any two nodes 
in \(s^{\prime}\), \ie one false positive and one false negative, in order to prevent the agent from taking the next 
optimal action \(a^{\prime}\) that has been learned so far (note that it may be different from the final optimal action). 
Specifically, for the DDQN agent, the attacker minimises \(\max_{a^{\prime}} Q(s^{\prime}+\delta, a^{\prime})\); 
for the A3C agent, the attacker minimises \(\max_{a^{\prime}} \pi(a^{\prime}|s^{\prime}+\delta)\). 

\begin{figure}[ht!]
\centering
\begin{subfigure}{.65\columnwidth}
  \centering
  \includegraphics[width=\columnwidth]{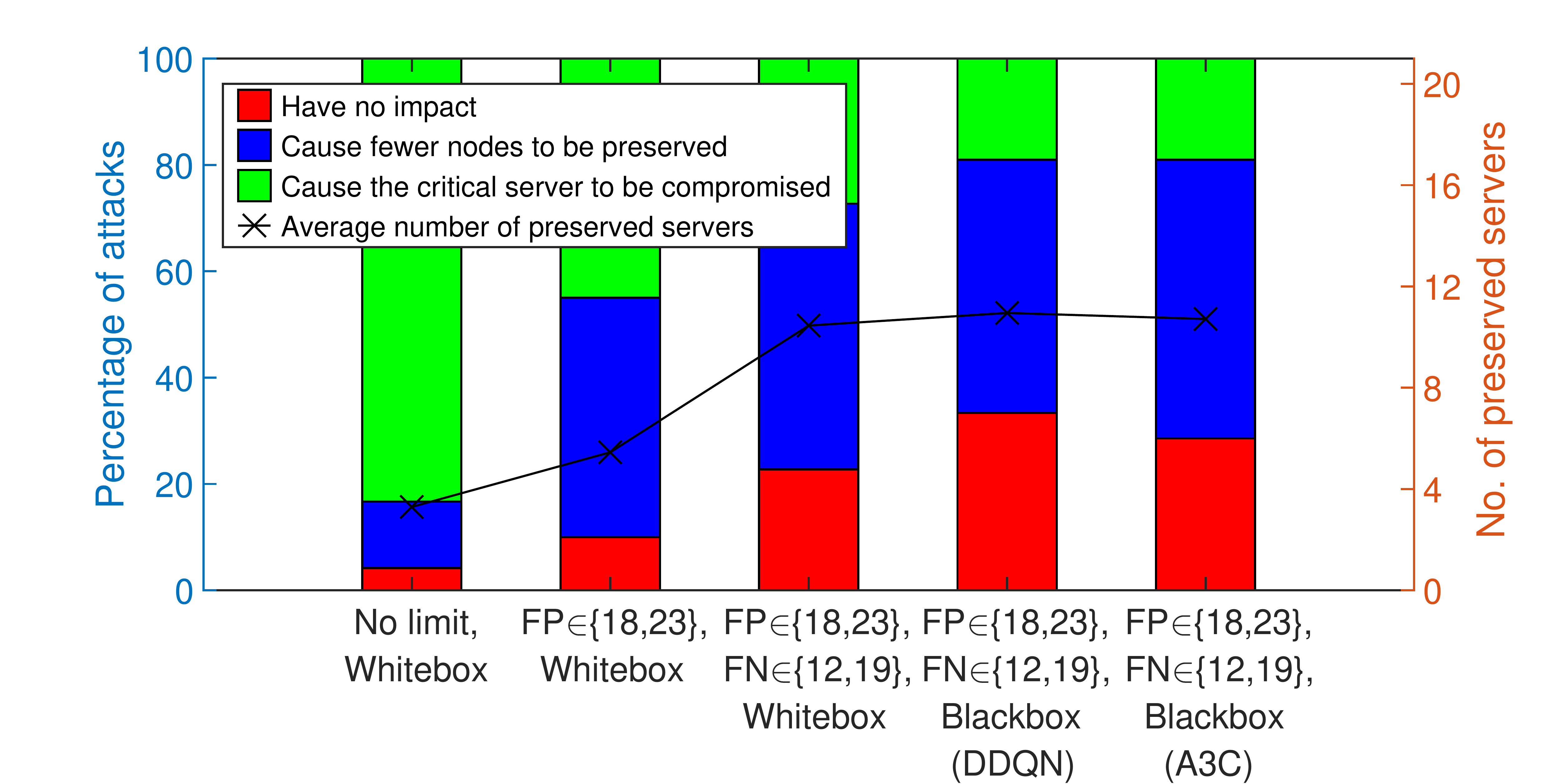}
  \caption{Indiscriminate attacks against DDQN}
  \label{figure_indiscriminate_ddqn}
\end{subfigure}
\begin{subfigure}{.65\columnwidth}
  \centering
  \includegraphics[width=\columnwidth]{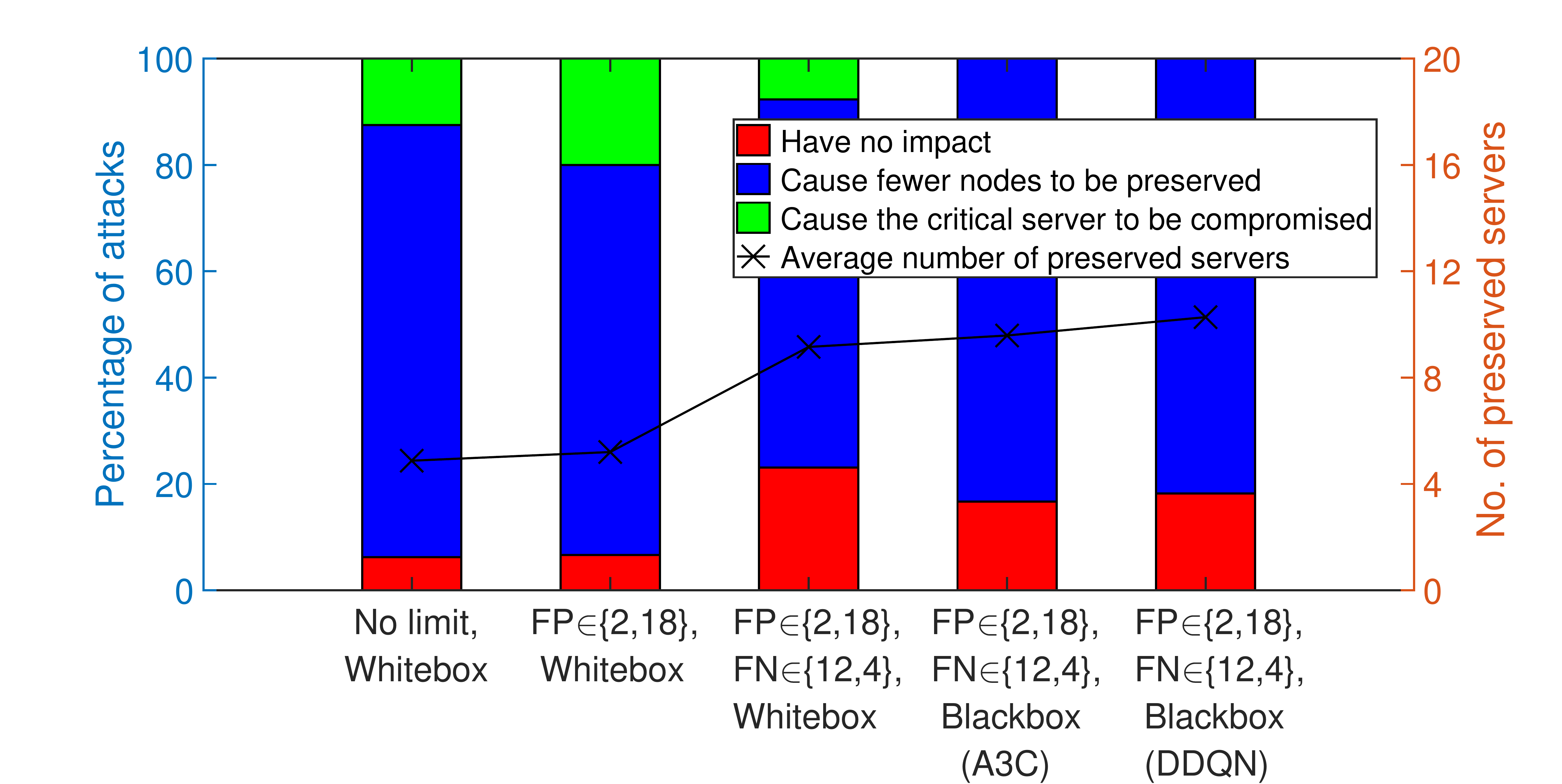}
  \caption{Indiscriminate attacks against A3C}
  \label{figure_indiscriminate_a3c}
\end{subfigure}
\caption{Indiscriminate attacks against DDQN \& A3C. The bars indicates the percentage of attacks (left \(y-\)axis) that 
(1) have no impact; (2) cause fewer nodes to be preserved; and (3) cause the critical server to be compromised. 
The line marked by ``\(\times\)'' indicates the average number of preserved servers (right \(y-\)axis). The five types of 
attacks are: (1) white-box, no limit on FNs\&FPs; (2) white-box, with limits on FP but not on FN, (3) white-box, with limits on 
both FP and FN; (4) black-box, same algorithm, with limits on both FPs and FNs; (5) black-box, different algorithm, with limits 
on both FPs and FNs.}
\label{figure_indiscriminate_attack}
\end{figure}

Fig.~\ref{figure_indiscriminate_attack} presents the results we obtained during our experiments. The leftmost bars in 
Figs.~\ref{figure_indiscriminate_ddqn} and ~\ref{figure_indiscriminate_a3c} suggest that the 
unlimited indiscriminate white-box attacks are very effective against both DDQN and A3C.
Specifically, the average number of preserved nodes decreases from 21 and 20 to 3.3 and 4.9, respectively.

\subsubsection{White-Box Attacks with Limits on the Choices of False Positive and False Negative}
\begin{figure}[ht!]
\centering
\begin{subfigure}{.5\columnwidth}
  \centering
  \includegraphics[width=\columnwidth]{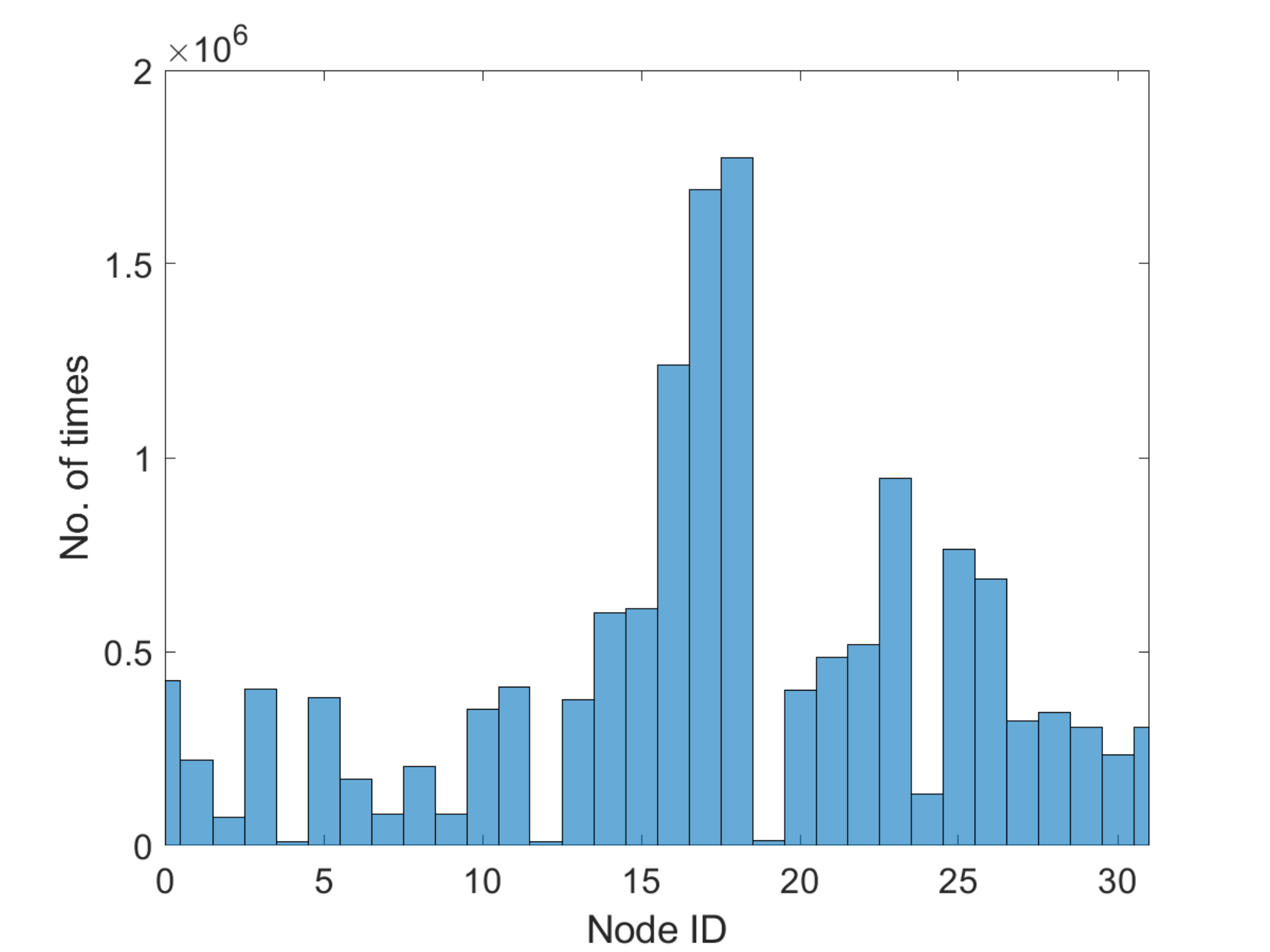}
  \caption{False positive}
  \label{figure_indiscriminate_dis_fp_ddqn}
\end{subfigure}%
\begin{subfigure}{.5\columnwidth}
  \centering
  \includegraphics[width=\columnwidth]{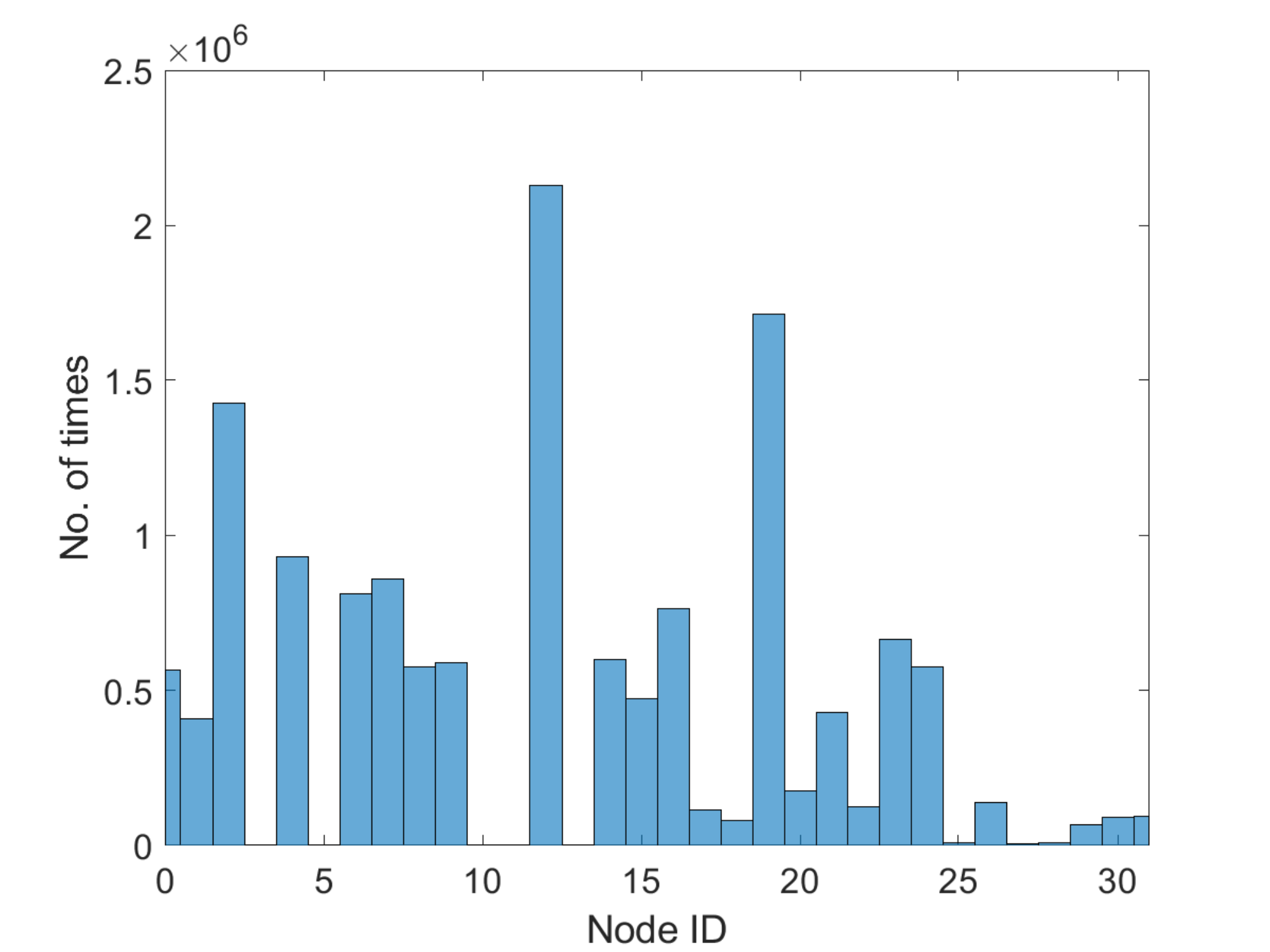}
  \caption{False negative}
  \label{figure_indiscriminate_dis_fn_ddqn}
\end{subfigure}
\caption{Histograms of the false positive and false negative nodes being selected against DDQN. 
N.B.: The node IDs \{0, 1, ..., 31\} are ordered and mapped to the node sequence 
\{1.1, 1.2, ..., 1.6, 2.1, ..., 2.8, 3.1, ..., 3.9, 4.1, ..., 4.9\}}
\label{figure_indiscriminate_dis_ddqn}
\end{figure}

As pointed out in Subsection~\ref{ps_casative_attacks}, even though the attacker only manipulates the states of two nodes 
each time, overall, they still need to be able to control a number of nodes, which is unlikely in practice. 
We calculate the number of times that each node is chosen in the above unlimited attacks, and find that some nodes are 
selected more frequently than others (Fig.~\ref{figure_indiscriminate_dis_ddqn}; 
the histograms for the A3C case are omitted due to similarity). 

Therefore, when poisoning the DDQN agent, we limit the false positive nodes to \{3.5 (node ID 18), 4.1 (node ID 23)\}, and limit 
the false negative nodes to \{2.7 (node ID 12), 3.6 (node ID 19)\}. We use node 4.1 instead of 3.4 (node ID 17), as otherwise 
both selected nodes would be from the target subnet and directly connected to the target, which is unlikely in real situations. 
In the A3C case, the false positive and false negative nodes are limited to \{1.3 (node ID 2), 3.5 (node ID 18)\}, 
and \{2.7 (node ID 12), 1.5 (node ID 4)\}, respectively.

The second and third bars in Figs.~\ref{figure_indiscriminate_ddqn} and~\ref{figure_indiscriminate_a3c} show that the limit has 
an obvious negative impact on the attack, especially the limit on the false negative nodes. Still, less than half of the nodes 
can be preserved on average, compared with the scenarios without attacks.

\subsubsection{Black-Box Attacks with Limits on the Choices of False Positive and False Negative Nodes}
In our black-box attacks (both intra- and cross-models), the attacker does not have access to the training model. 
Instead, they train their own agents first, and use the surrogate models to poison the training of the target models by 
choosing false positive and false negative nodes. 
Specifically, we have trained a few DDQN and A3C agents with a different number of hidden layers from the target model, 
and observed that these surrogates can still prevent the critical server from compromising.

As illustrated by the rightmost two bars in Figs.~\ref{figure_indiscriminate_ddqn} and~\ref{figure_indiscriminate_a3c}, 
black-box attacks are only slightly less effective than the counterpart white-box attacks despite the surrogate using 
a different model. This lends support that transferability also exists between RL algorithms, \ie attacks generated 
for one model may also transfer to another model.

\subsection{Attack II: Manipulate State---Targeted Attacks}
In the targeted attacks considered here, the attacker is assumed to know the sequence of final optimal actions, 
and attempts to minimise the probability of the agent following that sequence. It should be pointed out 
that we have also studied the case where the attacker instead maximises the probability of taking 
a specific non-optimal action for each step, but our results suggested that this is less effective.

We find that in targeted attacks, certain nodes are also selected more frequently as a false positive and false negative. 
In this scenario, we limit false positive nodes to (1) \{3.5 (node ID 18), 4.1 (node ID 23)\} against DDQN, 
(2) \{2.6 (node ID 11), 1.4 (node ID 3)\} against A3C, and limit false negative nodes to 
(1) \{1.5 (node ID 4), 2.1 (node ID 6)\} against DDQN, (2) \{4.1 (node ID 23), 2.4 (node ID 9)\} against A3C.

\begin{figure}[ht!]
\centering
\begin{subfigure}{.65\columnwidth}
  \centering
  \includegraphics[width=\columnwidth]{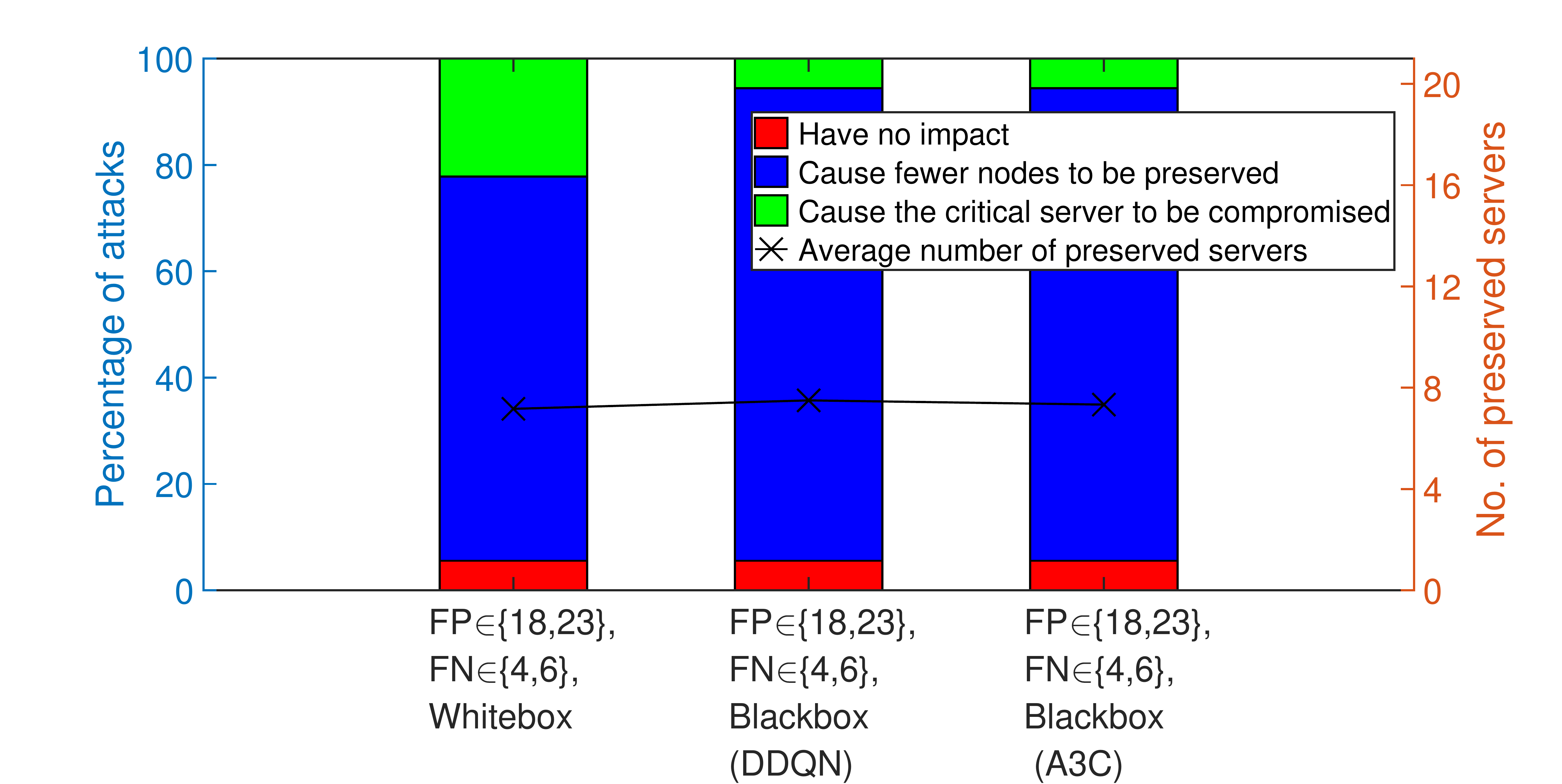}
  \caption{Targeted attacks against DDQN}
  \label{figure_targeted_ddqn}
\end{subfigure}
\begin{subfigure}{.65\columnwidth}
  \centering
  \includegraphics[width=\columnwidth]{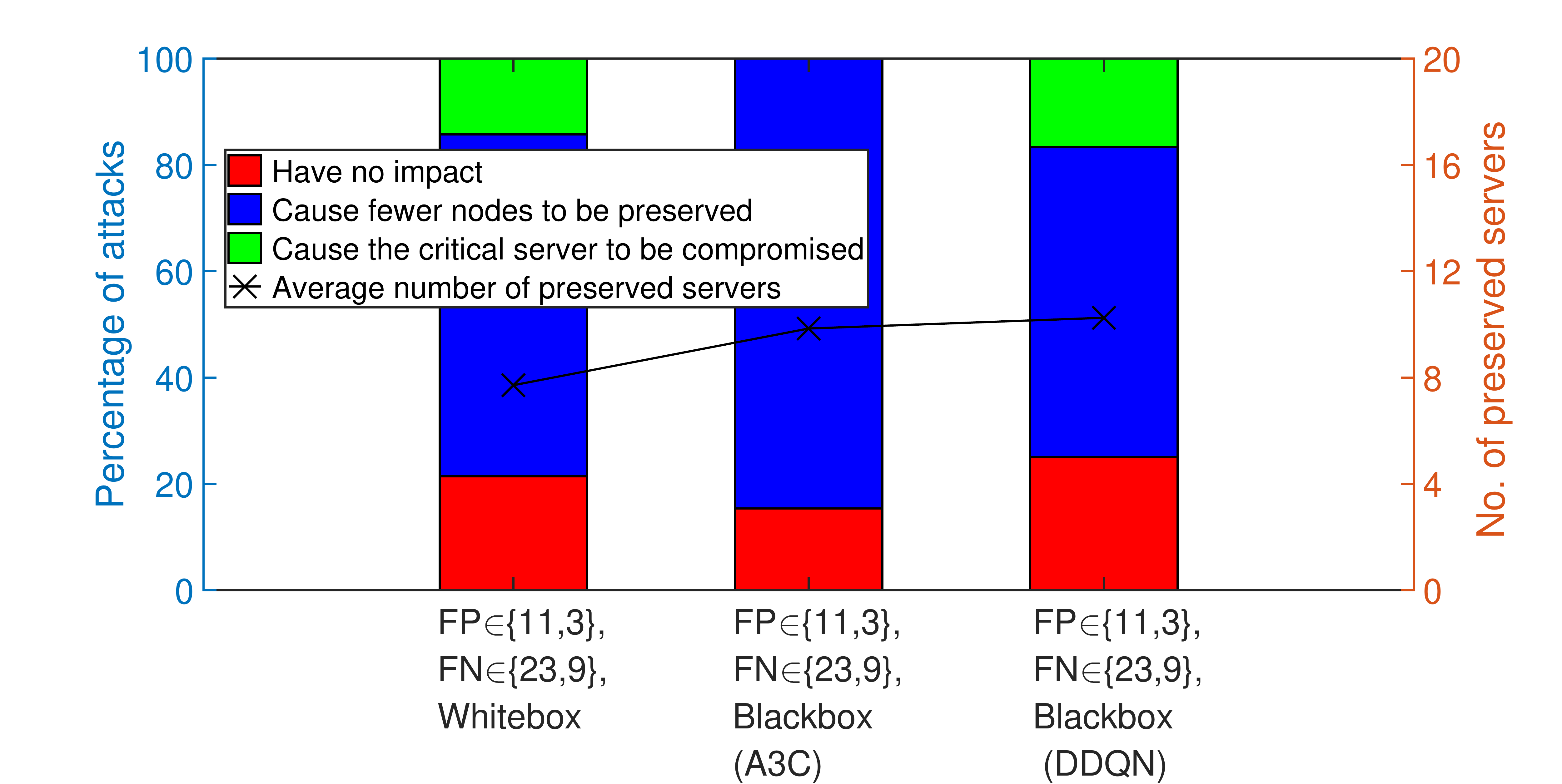}
  \caption{Targeted attacks against A3C}
  \label{figure_targeted_a3c}
\end{subfigure}
\caption{Targeted attacks against DDQN \& A3C. The bars indicate the percentage of attacks (left \(y-\)axis) that 
(1) have no impact; (2) cause fewer nodes to be preserved; and (3) cause the critical server to be compromised. 
The line marked by ``\(\times\)'' indicates the average number of preserved servers (right \(y-\)axis).}
\label{figure_targeted_attack}
\end{figure}

Our results, summarised in Fig.~\ref{figure_targeted_attack}, indicate that (1) compared with the results on indiscriminate attacks, 
targeted attacks work better, especially against DDQN (fewer nodes are preserved on average), as the attacker 
is more knowledgeable in this case; (2) similar to the indiscriminate case, black-box attacks achieve comparable results 
to the white-box attacks, further demonstrating the transferability between DDQN and A3C.

\subsection{Timing Limits for the Attacks}
The attacks discussed so far allowed the attacker to poison every experience obtained by the agent. A possible limitation on this
assumption is to examine whether these attacks can remain successful when the attacker can only manipulate part of the experiences. 
Therefore, in this subsection we shall look at attacks that poison only a subset (the first three steps) in each training episode. 

\begin{figure}[ht!]
\centering
\begin{subfigure}{.65\columnwidth}
  \centering
  \includegraphics[width=\columnwidth]{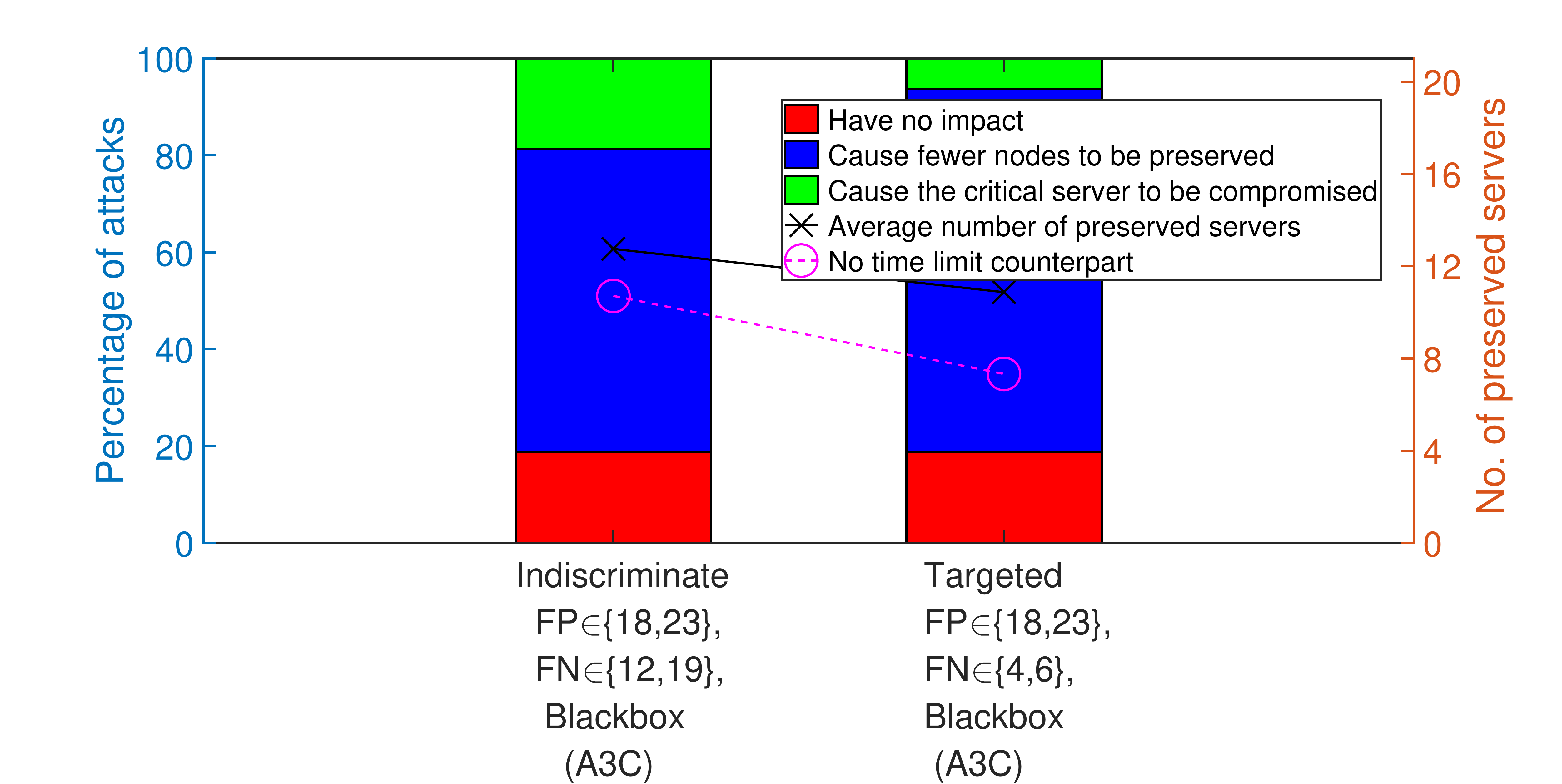}
  \caption{Time limited cross-model black-box attacks against DDQN}
  \label{figure_time_limit_ddqn}
\end{subfigure}
\begin{subfigure}{.65\columnwidth}
  \centering
  \includegraphics[width=\columnwidth]{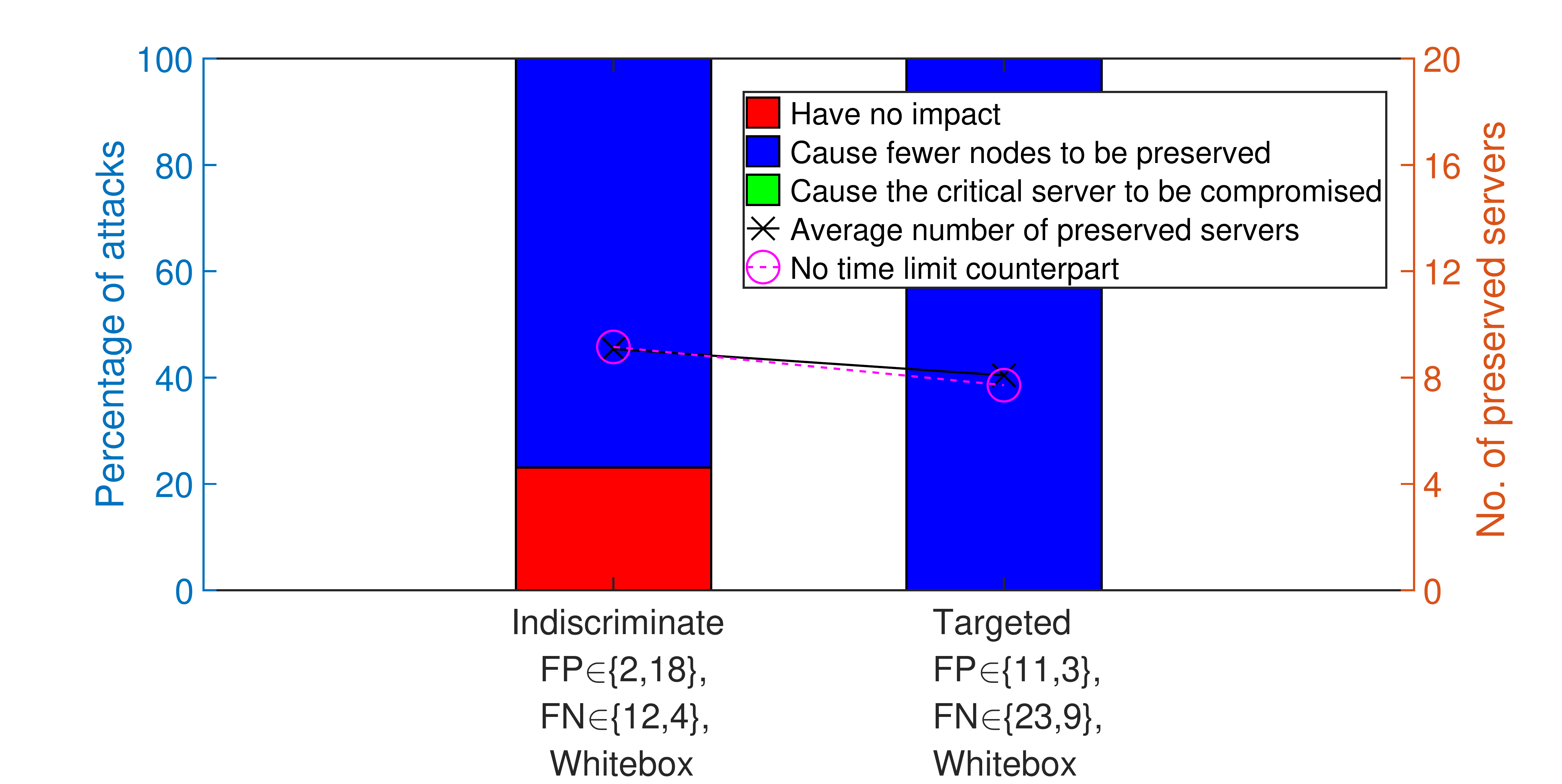}
  \caption{Time limited white-box attacks against A3C}
  \label{figure_time_limit_a3c}
\end{subfigure}
\caption{Attacks against DDQN\&A3C with time limit. The attacker only poisons the first three steps per training episode.}
\label{figure_time_limit}
\end{figure}

Figs.~\ref{figure_time_limit_ddqn} and~\ref{figure_time_limit_a3c} depict the results of the time-limited version of 
(cross-model) black-box attacks against DDQN and white-box attacks against A3C (both with limit on the choices of FPs and FNs), 
respectively. The results suggest that even though the time limit has a negative impact in every scenario studied, the attacks are 
still effective.

\subsection{Discussion on Countermeasures}
In supervised learning problems, adversarial training~\cite{szegedy_intriguing_2013,goodfellow_explaining_2014,tramer_ensemble_2017} 
has the defender
select a target point \((x, y)\) from the training set, 
modify \(x\) to \(x^{*}\) (\ie generates an adversarial sample), and then inject \((x^{*},y)\) back into the training set, 
under the implicit assumption that the true label \(y\) should not change given the instance has been only slightly perturbed.

In our RL setting, while the attacker manipulates the observed states to minimise 
the probability of the agent taking the optimal action \(a\) in state \(s\), the defender can construct adversarial samples that 
counteract the effect. For example, for each experience \((s, a, s^{\prime}, r)\), the defender can increase \(r\) by a small 
amount, \eg 5\% of the original value, given that \(r\) is positive and \(a\) is not chosen randomly (the probability of choosing 
an action randomly decreases as the training proceeds). The rationale behind this 
modification is that when the poisoning attack starts out, it is likely that \(a\) is still the optimal action (that has been 
learned so far) for state \(s\). If \(r\) is positive it means that  action \(a\) is a relatively good option for \(s\), and since 
the attacker has poisoned the state to prevent the agent from taking \(a\), we slightly increase \(r\) to encourage the agent to 
take action \(a\). 

\begin{figure}
\centering
\includegraphics[width=.65\textwidth]{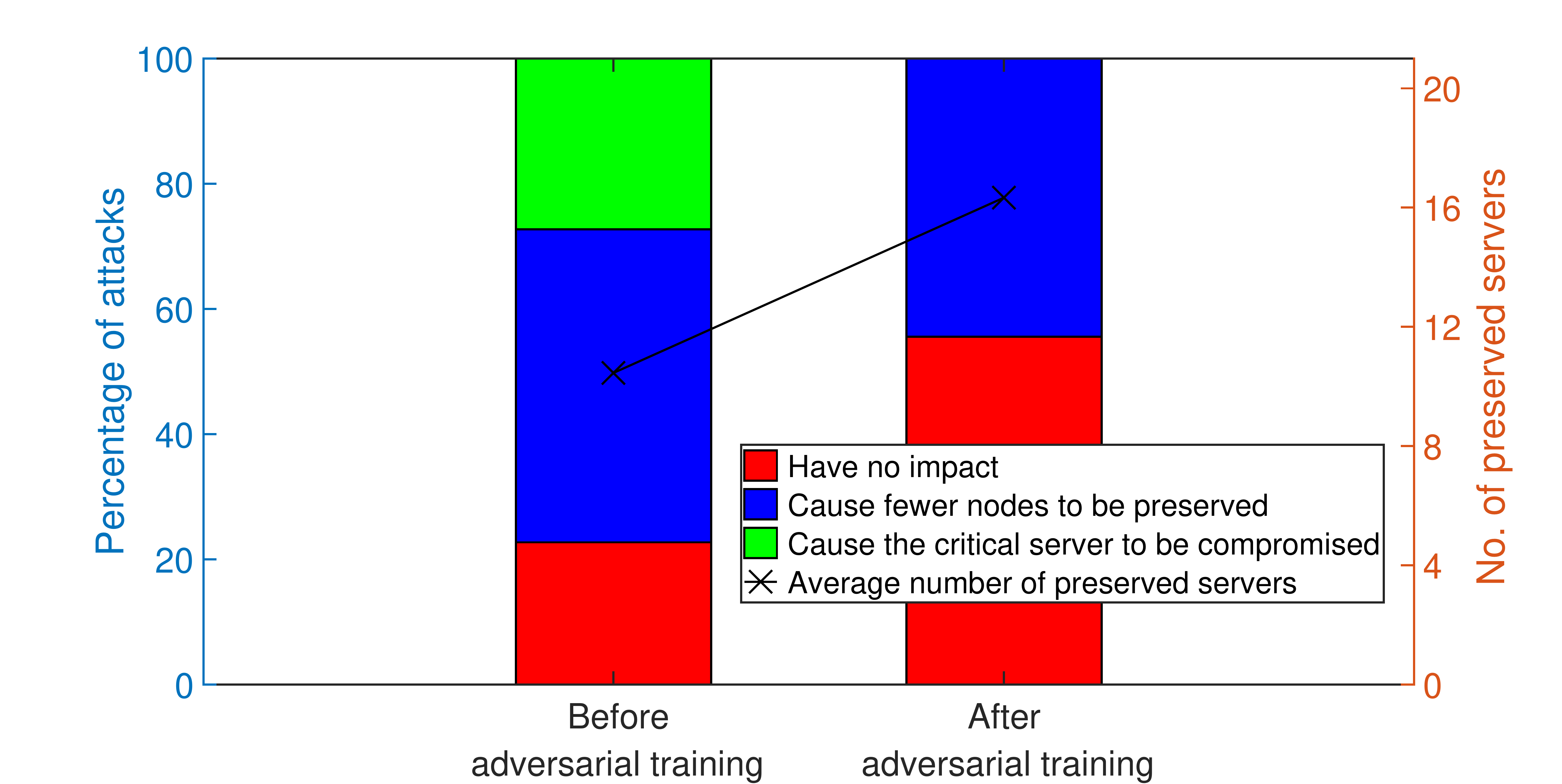}
\caption{Adversarial training against indiscriminate white-box attacks with limits on the choices of FPs and FNs (against DDQN), 
\(FP\in\{18,23\}, FN\in\{12,19\}\)}
\label{figure_adv_training}
\end{figure}

We have tested the above idea against indiscriminate white-box attacks with a limit on the choices of FPs and FNs against DDQN. 
Specifically, for each experience \((s, a, s^{\prime}, r)\) whose \(r\) is positive and \(a\) is not selected randomly, 
we change it to \((s, a, s^{\prime}, min(1.0, 1.05r))\). Note that our experimental results suggest that adding 5\% error to 
the reward signal when there is no attack will not prevent the agent from learning the optimal actions, although it may cause 
some delay. The results in Fig.~\ref{figure_adv_training} indicate that adversarial 
training can make the training process much less vulnerable. 

However, the results are still preliminary, and we plan to further investigate other forms of adversarial training. For example, 
Pinto \etal~\cite{pinto_robust_2017} model all potential disturbances as an extra adversarial agent, whose goal is to minimise the 
discounted reward of the leading agent. They formulate the policy learning problem as a two player zero-sum game, and propose an 
algorithm that optimises both agents by alternating learning one agent's policy with the other policy being fixed. In addition, we 
will also study the impact of the loss function, prioritised experience replay, ensemble adversarial training~\cite{tramer_ensemble_2017} 
and other, more intrusive types of attacks, where the adversary is aware of the defence method, and attacks the defended model.

\section{Related Work}\label{sec:related}
This section reviews ways in which attackers can target machine learning systems and current defence mechanisms. 
We first present a taxonomy on attacks against (primarily) supervised classifiers, and then summarise recent work that applies/modifies 
these attacks to manipulate RL systems. Finally, we review existing countermeasures against adversarial machine learning.

\subsection{Taxonomy of Attacks against Machine Learning Classifiers}
Barreno \etal \cite{barreno_security_2010} develop a qualitative taxonomy of attacks against machine learning classifiers 
based on three axes: influence, security violation and specificity.
\begin{itemize}
\item The axis of influence concerns the attacker's capabilities: in \emph{causative attacks}, 
the adversary can modify the training data to manipulate the learned model; in \emph{exploratory attacks}, 
the attacker does not poison training, but carefully alters target test instances to flip classifications. 
The resulting malicious instances which resemble legitimate data are called \emph{adversarial samples}.
\item The axis of security violation indicates the consequence desired by the attacker: \emph{integrity attacks} are an example of 
deception attacks (achieving uncertainty, incompleteness, etc.), resulting in indecision, delayed decisions, 
poor or even wrong decisions, in decision-making systems. In this type of attack, the malicious instances bypass the filter 
as false negatives. In contrast, \emph{availability attacks} seek to cause a denial-of-service against legitimate instances.
\item The axis of specificity refers to the target of the attacks: \emph{indiscriminate attacks} aim to degrade the classifier's 
performance overall, while \emph{targeted attacks} focus on a specific type of instance, or a specific instance.
\end{itemize}
Table~\ref{table_attacks_taxonomy} uses the taxonomy to classify previous work on adversarial machine learning against classifiers. 
As can be seen, more focus has been paid to exploratory integrity attacks. Presently, the Fast Gradient Sign Method (FGSM) attack
~\cite{goodfellow_explaining_2014} is  widely studied, and the C\&W attack~\cite{carlini_towards_2016} is the 
most effective found so far on the application domains tested, mostly in computer vision. 
Both of these attack methods can be used for targeted or indiscriminate attacks.

\begin{table}[t!]\small
\centering
\caption{Taxonomy of attacks on machine learners, with representative past work. 
As the taxonomy was designed for supervised learners, we include attacks on reinforcement learning in 
Section~\ref{sec:rl-attacks-lit}.}
\label{table_attacks_taxonomy}
\begin{tabular}{ l x{2.2cm} x{7.5cm} x{4.7cm} }
\hline
& & \textbf{Integrity} & \textbf{Availability}\\
\hline
\multirow{2}{*}{\textbf{Causative}} & \textbf{Targeted} (training set manipulation for specific errors) & 
Rubinstein \etal \cite{rubinstein_antidote:_2009}: boiling frog attacks against the PCA anomaly detection algorithm; \newline 
Li \etal \cite{li_data_2016}: poison training data against collaborative filtering systems; \newline 
Mei \& Zhu \cite{mei_using_2015}: identify the optimal training set to manipulate  different machine learners; \newline 
Burkard \& Lagesse \cite{burkard_analysis_2017}: targeted causative attack on Support Vector Machines that are learning 
from a data stream & Newsome \etal \cite{newsome_paragraph:_2006}: manipulate training set of classifiers for worms 
and spam to block legitimate instances; \newline Chung \& Mok \cite{chung_advanced_2007}: generate harmful signatures 
to filter out legitimate network traffic; \newline Nelson \etal \cite{nelson_exploiting_2008}: exploit statistical 
machine learning against a popular email spam filter\\
\cline{2-4}
& \textbf{Indiscriminate} (training set manipulation to maximise overall error rate) & Biggio \etal \cite{biggio_poisoning_2012}: 
inject crafted training data to increase SVM's test error; \newline 
Xiao \etal \cite{xiao_adversarial_2012}: label flips attack against SVMs; \newline 
Koh \& Liang \cite{koh_understanding_2017}: minimise the number of crafted training data via influence analysis & 
Newsome \etal \cite{newsome_paragraph:_2006}; Chung \& Mok \cite{chung_advanced_2007}; Nelson \etal \cite{nelson_exploiting_2008}\\
\hline
\multirow{2}{*}{\textbf{Exploratory}} & \textbf{Targeted} (test instance manipulation for specific errors) & 
Nelson \etal \cite{nelson_query_2012}: probe a classifier to determine good attack points; \newline 
Papernot \etal \cite{papernot_limitations_2016}: exploits forward derivatives to search for the minimum 
regions of the inputs to perturb; \newline Goodfellow \etal \cite{goodfellow_explaining_2014}: design the ``fast gradient 
sign method'' (FGSM) to generate adversarial samples; \newline 
Carlini \& Wagner \cite{carlini_towards_2016}: propose the C\&W method for creating adversarial samples; \newline 
Han \& Rubinstein \cite{han_adequacy_2017}: improve the gradient descent method by replacing with gradient quotient & 
Moore \etal \cite{moore_inferring_2006}: provide quantitative estimates of denial-of-service activity\\
\cline{2-4}
& \textbf{Indiscriminate} (test instance manipulation for misclassification) & Biggio \etal \cite{biggio_security_2014}: use 
gradient descent method to find attack instances against SVMs; \newline 
Szegedy \etal \cite{szegedy_intriguing_2013} demonstrate that changes imperceptible to human eyes can make 
DNNs misclassify an image; \newline Goodfellow \etal \cite{goodfellow_explaining_2014}; \newline 
Papernot \etal \cite{papernot_practical_2016, papernot_transferability_2016}: attack the target learner via a surrogate model; 
\newline Moosavi-Dezfooli \etal \cite{moosavi-dezfooli_universal_2016,moosavi-dezfooli_deepfool:_2016}: propose DeepFool 
that generates universal perturbations to fool multiple DNNs; \newline Carlini \& Wagner \cite{carlini_towards_2016}; \newline 
Nguyen \etal \cite{nguyen_deep_2015}: produce images that are unrecognisable to humans, but can be recognised by DNNs; \newline 
Han \& Rubinstein \cite{han_adequacy_2017} & Moore \etal \cite{moore_inferring_2006}\\
\hline
\end{tabular}
\end{table}

With further examination of the attacker's capabilities, in addition to the potential control over the training data, 
a powerful attacker may also know the internal architecture and parameters of the classifier. 
Therefore, a fourth dimension can be added to the above taxonomy according to attacker information: in \emph{white-box attacks}, 
the adversary generates malicious instances against the target classifier directly; while in \emph{black-box attacks}, 
since the attacker does not possess full knowledge about the model, they first approximate the target's model by training 
over a dataset from a mixture of samples obtained by observing the target's performance, and synthetically generating inputs 
and label pairs. Then if the reconstructed model generalises well, the crafted adversarial examples against this model can be 
transferred to the target network and induce misclassifications. 
Papernot \etal \cite{papernot_practical_2016,papernot_transferability_2016} have demonstrated the effectiveness of 
the black-box attack in certain specific domains. Specifically, they investigate intra- and cross-technique transferability 
between deep neural networks (DNNs), logistic regression, support vector machines (SVMs), decision trees and the $k$-nearest 
neighbour algorithm.

\subsection{Attacks against Reinforcement Learning}\label{sec:rl-attacks-lit}
In more recent studies, several papers have begun to study whether attacks against classifiers can also be 
applied to RL-based systems. Huang \etal \cite{huang_adversarial_2017} have shown that deep RL is vulnerable 
to adversarial samples generated by the Fast Gradient Sign Method~\cite{goodfellow_explaining_2014}. 
Their experimental results demonstrate that both white-box and black-box attacks are effective, 
even though the less knowledge the adversary has, the less effective the adversarial samples are.

Behzadan \& Munir \cite{behzadan_vulnerability_2017} establish that adversaries can interfere with the training process of 
DQNs, preventing the victim from learning the correct policy. 
Specifically, the attacker applies minimum perturbation to the state observed by the target, 
so that a different action is chosen as the optimal action at the next state. The perturbation is generated 
using the same techniques proposed against DNN classifiers. In addition, the authors  demonstrate 
the possibility of policy manipulation, where the victim ends up with choosing the actions selected by the adversarial policy.

Lin \etal \cite{lin_tactics_2017} propose two kinds of attacks against deep reinforcement learning agents. 
In \emph{strategically-timed attacks}, instead of crafting the state at each time step, the adversary identifies a 
subset of most vulnerable steps, and uses the C\&W attack~\cite{carlini_towards_2016} to perturb the corresponding states. 
In \emph{enchanting attacks}, the adversary uses sampling to iteratively find a sequence of actions that will 
take the agent to the target state, and craft the current state so that the agent will follow the next required action.

\subsection{Adversarial Machine Learning Defences}
A number of countermeasures have been proposed since the discovery of adversarial samples. These can be roughly categorised into 
two classes: \emph{data-driven defences} and \emph{learner robustification}. 

\subsubsection{Data-driven Defences}
This class of defences are data driven -- they either filter out the malicious data, inject adversarial samples into the 
training dataset, or manipulate features via projection. These approaches are akin to black-box defences since they make little 
to no use of the learner.

\begin{itemize}
\item Filtering instances. These counter-measures assume that the poisoning data in the training dataset or the 
adversarial samples against the test dataset either exhibit different statistical features, or follow a different distribution. 
Therefore, they propose to identify and filter out the injected/perturbed data. For example, 
Laishram and Phoha~\cite{laishram_curie:_2016} design a method called Curie to protect SVM classifiers against poisoning attacks that 
flip labels in the training dataset. Steinhardt, Koh and Liang~\cite{steinhardt_certified_2017} study the worst-case loss of 
both data dependent and independent sanitisation methods. Metzen \etal \cite{metzen_detecting_2017} propose to train a 
detector network that takes input from intermediate layers of a classification network, and filters out adversarial samples. 
Feinman \etal \cite{feinman_detecting_2017} use (i) kernel density estimates in the feature space of a final hidden layer, 
and (ii) Bayesian neural network uncertainty estimates to detect the adversarial samples against DNNs. 
Li \etal \cite{li_adversarial_2016} apply principal component analysis (PCA) on the convolutional layers of an original 
convolutional neural network (CNN), and use the extracted statistics to train a cascade classifier that can distinguish 
between valid and adversarial data.

\item Injecting data. Goodfellow \etal \cite{goodfellow_explaining_2014} attribute the existence of adversarial samples 
to the ``blind spots'' of the training algorithm, and propose injecting adversarial examples into training, in order to improve 
the generalisation capabilities of DNNs~\cite{szegedy_intriguing_2013,goodfellow_explaining_2014} -- akin to active learning 
in non-adversarial settings. Tramer \etal \cite{tramer_ensemble_2017} extend such adversarial training methods by incorporating 
perturbations generated against other models.

\item Projecting data. Previous work has shown that high dimensionality facilitates the generation of adversarial 
samples, resulting in an increased attack surface. For example, Wang, Gao and Qi~\cite{wang_theoretical_2016} theorise that a single 
unnecessary feature can ruin the robustness of a model. To counter this, data can be projected into lower-dimensional 
space before testing. Specifically, Bhagoji \etal \cite{bhagoji_dimensionality_2017} and 
Zhang \etal \cite{zhang_adversarial_2016} propose defence methods based on dimensionality reduction via principal component 
analysis (PCA), and reduced feature sets, respectively. Das \etal \cite{das_keeping_2017} demonstrate that JPEG compression 
can be used as a pre-processing step to defend against evasion attacks in computer vision, potentially because JPEG 
compression removes high-frequency signal components, which helps remove imperceptible perturbations. However, these results 
contradict those obtained by Li and Vorobeychik \cite{li_feature_2014}, which suggests that more features should be used when facing 
adversarial evasion.
\end{itemize}

\subsubsection{Learner Robustification}
Rather than focusing solely on training and test data, this class of methods---which are white-box in nature---aim to design 
models to be less susceptible to adversarial samples in the first place.

\begin{itemize}
\item Stabilisation. Zheng \etal \cite{zheng_improving_2016} design stability training that modifies the model's 
objective function by adding a stability term. Their experimental results demonstrate that such a modification stabilises 
DNNs against small perturbations in the inputs. Papernot \etal \cite{papernot_distillation_2015} provide further examples 
using a distillation strategy against a saliency-map attack. However, this method has been shown to be ineffective by 
Carlini and Wagner \cite{carlini_defensive_2016}. Hosseini \etal \cite{hosseini_blocking_2017} propose to improve adversarial 
training by adding an additional ``NULL'' class, and attempt to classify all adversarial samples as invalid.

\item Moving target. Sengupta \etal \cite{sengupta_securing_2017} apply moving target defences against exploratory 
attacks: instead of using a single model for classification, the defender prepares a pool of models, and for each image 
to be classified, one trained DNN is picked following some specific strategy. The authors formulate the interaction as a 
Repeated Bayesian Stackelberg Game, and show that their approach can decrease the attack's success rate, while maintaining 
high accuracy for legitimate users.

\item Robust statistics. Another avenue that has remained relatively unexplored is to leverage ideas from robust 
statistics, such as influence functions, break-down points, and $M$-estimators with robust loss functions (such as the Huber loss) 
that place diminishing cost to increasingly erroneous predictions. Rubinstein \etal \cite{rubinstein_antidote:_2009} 
were the first to leverage robust statistics in adversarial learning settings for cyber-security, by applying a robust 
form of principal components analysis that optimises median absolute deviations instead of variance to defend against causative 
attacks on network-wide volume anomaly detection. Recently, interest in the theoretical computer science community 
has turned to robust estimation in high dimensions, \eg~ Diakonikolas \etal \cite{diakonikolas_robust_2016}.
\end{itemize}

\subsubsection{Lessons Learned}
Despite many defences proposed, several recent studies \cite{he_adversarial_2017,carlini_adversarial_2017} point out that 
most of these methods (i) unrealistically assume that the attacker is not aware of the defence mechanism, and (ii) only 
consider relatively weak attacks, \eg FGSM~\cite{goodfellow_explaining_2014}. Negative results are reported on the effectiveness 
of these methods against adaptive attackers that are aware of the defence and act accordingly, and against the 
C\&W attack~\cite{carlini_towards_2016} (empirically the most efficient exploratory attack proposed so far).

Specifically, He \etal \cite{he_adversarial_2017} demonstrate that the following recently proposed methods cannot defend 
against adaptive adversaries: (i) feature squeezing \cite{xu_feature_2017}, (ii) specialists+1 ensemble method 
\cite{abbasi_robustness_2017}, and (iii) ensemble of methods in the papers 
\cite{metzen_detecting_2017,feinman_detecting_2017,gong_adversarial_2017}. 
Carlini and Wagner \cite{carlini_adversarial_2017} show that ten detection methods for adversarial samples can be defeated, 
either by the C\&W attack \cite{carlini_towards_2016}, or white-box/black-box attacks. The authors conclude that the most 
effective defence so far for DNNs is to apply randomness to the DNN, because it makes generating adversarial samples against 
the target classifier as difficult as generating transferable adversarial samples. More recently, Athalye \etal 
\cite{athalye_obfuscated_2018} show that defences relied on obfuscated gradients can also be circumvented.

While most of the previous work on adversarial machine learning focused on the vision domain, this paper applies 
RL in cyber security, and studies how it reacts against different forms of causative integrity attacks.

\section{Conclusions and Future Work}\label{sec:conc}
In this paper, we demonstrated the feasibility of developing autonomous defence in SDN using RL algorithms. 
In particular, we studied the impact of different forms of causative attacks, and showed that even though these attacks  
might cause RL agents to take sub-optimal actions, adversarial training could be applied to mitigate the impact.

For future work, we plan to (1) use a traffic generator to introduce background traffic between nodes, and use network performance 
metrics to replace the current binary states; (2) consider different types of network traffic, so that the actions of the RL agent 
could include partial isolation in terms of blocking certain protocols between nodes; (3) change full observability of the network 
status to partial observability---the defender may have limited resources, and the attacker may not know the entire topology; 
and (4) remove limiting assumptions, \eg the attacker having to compromise all nodes along the path to the critical server.

\bibliographystyle{splncs04}
\bibliography{references}

\end{document}